\documentclass{article}

\usepackage{amssymb}
\usepackage{fleqn}
\usepackage{epsfig}
\usepackage{graphicx}
\usepackage{amsmath}

\def\beq{\begin{equation}}
\def\eeq{\end{equation}}
\def\bea{\begin{eqnarray}}
\def\eea{\end{eqnarray}}
\def\bq{\begin{quote}}
\def\eq{\end{quote}}

\def\gappeq{\mathrel{\rlap {\raise.5ex\hbox{$>$}}{\lower.5ex\hbox{$\sim$}}}}
\def\lappeq{\mathrel{\rlap{\raise.5ex\hbox{$<$}}{\lower.5ex\hbox{$\sim$}}}}
\def\Toprel#1\over#2{\mathrel{\mathop{#2}\limits^{#1}}}

\begin{document}
\allowdisplaybreaks

\pagestyle{empty}
\begin{flushright}
ROME1/1415/05~

DSFNA1/35/2005
\end{flushright}
\vspace*{15mm}

\begin{center}
\textbf{THRESHOLD RESUMMED SPECTRA IN $B\rightarrow X_u l \nu$ DECAYS IN NLO (III)} \\[0pt]

\vspace*{1cm}

\textbf{Ugo Aglietti}\footnote{e-mail address: Ugo.Aglietti@roma1.infn.it} \\[0pt]

\vspace{0.3cm}
Dipartimento di Fisica,\\
Universit\`a di Roma ``La Sapienza'', \\
and I.N.F.N.,
Sezione di Roma, Italy. \\[1pt]

\vspace{0.3cm}
\textbf{Giulia Ricciardi}\footnote{e-mail address:
Giulia.Ricciardi@na.infn.it} \\ [0pt]

\vspace{0.3cm} Dipartimento di Scienze Fisiche,\\
Universit\`a di Napoli ``Federico II'' \\
and I.N.F.N.,
Sezione di Napoli, Italy. \\ [1pt]

\vspace{0.3cm}
\textbf{Giancarlo Ferrera}\footnote{e-mail address: Giancarlo.Ferrera@roma1.infn.it} \\[0pt]

\vspace{0.3cm}
Dipartimento di Fisica,\\
Universit\`a di Roma ``La Sapienza'', \\
and I.N.F.N.,
Sezione di Roma, Italy. \\[1pt]

%$~~~$ \\[0pt]
\vspace*{1cm} \textbf{Abstract} \\[0pt]
\end{center}
We resum to next-to-leading order (NLO) the distribution in the
light-cone momentum $p_+ = E_X - |\vec{p}_X|$ and the spectrum in
the electron energy $E_e$ in the semileptonic decays $B\rightarrow
X_u l \nu$, where $E_X$ and $\vec{p}_X$ are the total energy and
three-momentum of the final hadron state $X_u$ respectively. 
By expanding our formulae, we obtain the coefficients of all the infrared
logarithms $\alpha_S^n \, L^k$ at $O(\alpha_S^2)$ and at $O(\alpha_S^3)$,  
with the exception of the $\alpha_S^3 \, L$ coefficient.
By comparing our $O(\alpha_S^2 \, n_f)$ result for the $p_+$ distribution
with a recent Feynman diagram computation, we obtain 
an explicit and non trivial verification of our resummation scheme at 
the two-loop level ($n_f$ is the number of massless quark flavors).
We also discuss the validity of the so-called Brodsky-Lepage-Mackenzie (BLM) 
scheme in the evaluation of the semileptonic spectra by comparing it
with our results, finding that it is a reasonable approximation.
Finally, we show that the long-distance phenomena, such as Fermi motion,
are expected to have a smaller effect in the electron spectrum than in the 
hadron mass distribution.

\vspace*{2cm} 
\noindent

\vfill\eject

\setcounter{page}{1} \pagestyle{plain}

\section{Introduction}

The structure of final hadronic states in semileptonic $B$ decays
\begin{equation}
\label{startsl}
B \, \rightarrow \, X_u \, + \, l \, + \, \nu,
\end{equation}
where $X_u$ is any hadronic final state coming from the fragmentation of the $up$
quark and $l+\nu$ is a lepton-neutrino pair,
is usually studied by measuring different spectra, such as the charged-lepton
energy distribution, the dilepton invariant mass spectrum, the hadron mass
distribution, etc.
The computation of these spectra is often rather involved
because of the occurrence of different long-distance effects, both perturbative
and non-perturbative, such as large logarithms, Fermi motion, hadronization, etc.
These effects are often substantial in the end-point regions relevant
for the experimental analysis.

Let us summarize in physical terms the main properties of
semi-inclusive $B$ decays \cite{me,noi,noi2}. We begin by
considering the radiative decays
\begin{equation}
\label{startrd}
B \, \rightarrow \, X_s \, + \, \gamma
\end{equation}
which have a simpler long-distance structure than the semileptonic decays (\ref{startsl}).
That is because, in the radiative case, the tree-level process is the two-body decay
\begin{equation}
b \, \rightarrow \, s \, + \, \gamma
\end{equation}
having the large final hadron energy
\begin{equation}
\label{bornenergy}
2 \, E_X \, = \, m_b \, \left(1 \, - \, \frac{m_s^2}{m_b^2}\right) \, \cong \, m_b,
\end{equation}
where in the last member we have neglected the small (compared with the beauty mass $m_b$)
strange mass $m_s$.
The final hadronic energy $E_X$ fixes the hard scale $Q$ in the decay:
\begin{equation}
\label{adnauseam}
Q \, = \, 2\, E_X,
\end{equation}
so that $Q \cong m_b$.
We are interested in the threshold region
\begin{equation}
\label{closetoborn}
m_X \, \ll \, E_X,
\end{equation}
which can be considered a kind of ``perturbation'' of the tree-level process
due to soft-gluon effects, where $m_X$ is the final hadron mass.
The hadron mass $m_X$, which vanishes in lowest order, remains indeed small in higher
orders because of the condition (\ref{closetoborn}), while the large tree-level
hadron energy (\ref{bornenergy}) is only mildly increased by soft emissions.
The perturbative expansion is therefore controlled by the coupling
\begin{equation}
\label{smallcoup}
\alpha\left(Q\right) \, = \, \alpha\left(m_b\right) \, \cong \, 0.22 \, \ll \, 1,
\end{equation}
where $\alpha \, = \, \alpha_S$ is the QCD coupling, and it is therefore
legitimate.
 
In semileptonic decays, the tree-level process is instead the {\it three-body} 
decay
\begin{equation}
b \, \rightarrow \, u \, + \, l \, + \, \nu
\end{equation}
and the hadron energy $E_X$ (i.e. the energy of the $up$ quark) can become
substantially smaller than half of the beauty mass $m_b/2$. 
In other words, kinematical configurations are possible with
\beq
\label{comerad}
E_X \, \approx \, \frac{m_b}{2}
\eeq
as well as with
\footnote{
It is clear that the condition
\beq
E_X \, \gg \, \Lambda
\eeq
must always be verified in order to deal with a hard process and 
to be justified in the use of perturbation theory.
Since the beauty mass $m_b$ is larger 
than the hadronic scale $\Lambda$ by an order of
magnitude only, it is in practise not easy to satisfy all the conditions
above. In any case, the theory is constructed by taking the limit
$m_b \, \to \, \infty$.
}
\beq
\label{noncomerad}
E_X \, \ll \, m_b.
\eeq
Consider for example the kinematical configuration, 
in the $b$ rest frame, with the electron and the neutrino
parallel to each other, for which condition (\ref{comerad}) holds
true, or the configuration with the leptons back to back, 
each one with an energy $\approx \, m_b/2$,
for which condition (\ref{noncomerad}) is instead verified.  
This fact is the basic additional complication in going from the radiative 
decays to the semileptonic ones: the hard scale is no more fixed by the
heavy flavor mass but it depends on the kinematics according to
eq.~(\ref{adnauseam}).

Before going on, let us give the definition of a short-distance quantity 
in perturbation theory.
Let us consider a process characterized by a hard scale 
\beq
Q \, \gg \, \Lambda,
\eeq
where $\Lambda$ is the QCD scale, and by the infrared scales 
\beq
\frac{m_X^4}{Q^2}, ~~ m_X^2  \,\, \ll \,\, Q^2
\eeq  
--- the soft scale and the collinear scale respectively.
If infrared effects are absent in the quantity under consideration, 
logarithmic terms of the form 
\begin{equation}
\label{infraredsi}
~~~~~~~~~~~~~~~~~~~~~~~~~~~~~~~~~~~~~~~~~~~~~~~~~~~
\alpha^n \, \log^k \, \frac{Q^2}{m_{X}^{\,2}}
~~~~~~~~~~~~~~~~~~(n \, = \, 1,2, \, \cdots \, \infty,
~ \, k \, = \, 1,2, \, \cdots \, 2n)
\end{equation}
must not appear in its perturbative expansion.
The presence of terms of the form (\ref{infraredsi}) would indeed signal 
significant contributions from small momentum scales.
That is because these terms originate from the integration of the infrared-enhanced 
pieces of the $QCD$ matrix elements from the hard scale down to one of the 
infrared scales:
\beq
\label{integrali}
\int_{m_X^4/Q^2}^{Q^2} \, \frac{dk_{\perp}^2}{k_{\perp}^2},
~~~~~
\int_{m_X^2}^{Q^2} \, \frac{dk_{\perp}^2}{k_{\perp}^2}
~~~~~ \Rightarrow ~~~~~
\log \, \frac{Q^2}{m_X^2}.
\eeq
If perturbation theory shows a significant contribution from small
momentum scales to some cross section or decay width, we believe there
is no reason to think that the same should not occur in a non-perturbative 
computation.

On the contrary, a quantity such as a spectrum or a ratio of spectra
is long-distance dominated if it contains infrared logarithmic terms in the perturbative 
expansion of the form (\ref{infraredsi}).
These terms must be resummed to all orders of perturbation theory in the threshold region 
(\ref{closetoborn}).
For hadronic masses as small as
\begin{equation}
m_X \, \approx \, \sqrt{\Lambda \, Q},
\end{equation}
the non-perturbative effects related to the soft interations of the $b$ quark
in the $B$ meson (the so-called Fermi motion) also come into play and have to be 
included by means of a structure function
\footnote{
The first and the second integral on the l.h.s. of eq.~(\ref{integrali})
are related to the structure function and to its coefficient
function respectively.
}.
Our criterion to establish whether a quantity is short-distance or it is not
is rather ``narrow''; we believe that it is a fundamental one and we use it sistematically,
i.e. we derive all its consequences. The consequences, as we are going to show, 
are in some cases not trivial.

Factorization and resummation of threshold logarithms in the radiative 
case is similar to that of shape variables in $e^+e^-$ annihilations to hadrons.
One can write:
\beq
\label{singola}
\frac{1}{\Gamma_r} \int_0^{t_s} \frac{d\Gamma_r}{d t_s'} \, dt_s' 
\,=\, C_r\left[\alpha(m_b)\right]\,
\Sigma\left[t_s;\, \alpha(m_b)\right] \, + \, D_r\left[t_s;\, \alpha(m_b)\right],
\eeq
where $\Gamma_r$ is the inclusive radiative width,
we have defined
\beq
~~~~~~~~~~~~~~~~~~~~~~~~~~~~~~~~~~~~~~~~~~~~~~~~
t_s \, \equiv \, \frac{m_{X_s}^2}{m_b^2}
~~~~~~~~~~~~~~~~~~~~~~~~~~~~~~~~~~(0 \, \le \, t_s \, \le \, 1)
\eeq
and:
\begin{itemize}
\item
$C_r\left( \alpha \right)$ is a short-distance, process-dependent 
coefficient function, independent on the hadron variable $t_s$
and having an expansion in powers of $\alpha$:
\beq
C_r\left( \alpha \right) \, = \,
1 \, + \, \sum_{n=1}^{\infty} C_{\,r}^{(n)} \, \alpha^n.
\eeq
The explicit expression of the first-order correction 
$C_{\, r}^{(1)}$ has been given in \cite{me,acg};
\item
$\Sigma(u;\,\alpha)$ is the universal QCD form factor for heavy 
flavor decays and has a double expansion of the form:
\bea
\label{Sigma}
\Sigma[u;\,\alpha] & = & 
1 \, + \, \sum_{n=1}^{\infty} \sum_{k=1}^{2n} \, 
\Sigma_{nk} \, \alpha^n \, \log^k \frac{1}{u} \, =
\nonumber
\\
& = & 1 
\, - \, \frac{1}{2}\,\frac{\alpha\,C_F}{\pi} \, \log^2 \frac{1}{u}
\, + \, \frac{7}{4}\, \frac{\alpha\,C_F}{\pi} \, \log \frac{1}{u}
\, + \, \frac{1}{8}\,\left(\frac{\alpha\,C_F}{\pi}\right)^2 \, \log^4 \frac{1}{u}
\, + \, \cdots,
\eea
where $C_F \, = \, (N_C^2-1)/(2N_C)$ is the Casimir of the 
fundamental representation of the group $SU(3)$
and $N_C \, = \, 3$ is the number of colors.
In higher orders, as it is well known, $\Sigma$ contains at most 
two logarithms for each power of $\alpha$, coming from the overlap 
of the soft and the collinear region in each emission;
\item
$D_r\left(t_s; \, \alpha \right)$ is a short-distance-dominated,
process-dependent remainder function, not containing infrared logarithms 
and vanishing for 
$t_s \, \rightarrow \, 0$ as well as for $\alpha \, \rightarrow \, 0$:
\beq
D_r\left(t_s; \, \alpha \right) \, = \,
\sum_{n=1}^{\infty} D_{\,r}^{(n)}(t_s) \, \alpha^n.
\eeq
The first-order correction $D_{\,r}^{(1)}(t_s)$ has been explicitly 
given in \cite{me,acg}.
\end{itemize}

\noindent
Factorization and resummation of threshold logarithms in the semileptonic 
case is conveniently made starting with distributions not integrated over
$E_X$, i.e. not integrated over the hard scale $Q$.
The most general distribution in process (\ref{startsl})
is a triple distribution, which has a resummed expression
of the form \cite{me,noi,noi2}:
\begin{equation}
\label{tripla}
\frac{1}{\Gamma} \int_0^u \frac{d^3\Gamma}{dx dw du'} \, du' 
\, = \, C\left[x,w;\alpha(w\,m_b)\right]\,
\Sigma\left[u;\alpha(w\,m_b)\right] \, + \, D\left[x,u,w;\alpha(w\,m_b)\right],
\end{equation}
where we have defined the following kinematical variables:
\begin{equation}
x \, = \, \frac{2 E_l}{m_b}~~(0\le x \le 1);~~~~~
w \, = \, \frac{Q}{m_b}~~(0\le w \le 2);~~~~~
u \, = \, \frac{1 - \sqrt{1 - \left(2m_X/Q\right)^2} }{1 + \sqrt{1 - \left(2m_X/Q\right)^2} }
\, \simeq \, \left(\frac{m_X}{Q}\right)^2
~~(0\le u \le 1).
\end{equation}
In the last member we have kept the leading term in the threshold region $m_X \ll Q$ only.
$Q$, the hard scale, is given by eq.~(\ref{adnauseam}).
$\Gamma$ is the total semileptonic width,
\begin{equation}
\label{K1}
\Gamma \, = \, \Gamma_0\left[1 
\, + \, \frac{\alpha \, C_F}{\pi}
\left( \frac{25}{8} - \frac{\pi^2}{2} \right) \, + \, O(\alpha^2) \right],
\end{equation}
where $\Gamma_0 \, = \, G_F^2 m_b^5  |V_{ub}|^2 /(192\pi^3)$ is the tree-level width 
and $m_b$ is the pole mass.
Eq.~(\ref{tripla}) is a ``kinematical''
generalization of eq.~(\ref{singola}) together with the quantities involved, 
which all have an expansion in powers of $\alpha$:
\begin{itemize}
\item
$C\left[x,w;\alpha\right]$, a short-distance, process-dependent 
coefficient function, dependent on the hadron and lepton energies
but independent on the hadron variable $u$.
The explicit expressions of the first two orders have been given 
in \cite{me,noi};
\item
$\Sigma[u;\,\alpha]$, the universal $QCD$ form factor for heavy 
flavor decays, which now is evaluated for a coupling with a general
argument $Q \, = \, w \, m_b$;
\item
$D\left[x,u,w;\alpha\right]$, a  short-distance dominated,
process-dependent remainder function, depending on all the kinematical variables,
not containing infrared logarithms and vanishing for $u\rightarrow 0$ 
as well as for $\alpha\rightarrow 0$.
The first-order correction has been explicitly given in \cite{noi}.
\end{itemize}
The resummed formula on the r.h.s. of eq.~(\ref{tripla}) 
has been derived with general arguments in an effective field theory 
which hold to any order in $\alpha$; 
one basically considers the infinite-mass limit for the beauty quark
\beq
m_b \, \to \, \infty,
\eeq
while keeping the hadronic energy $E_X$ and the hadronic mass $m_X$ fixed.
The main result is that the hard scale $Q$ enters:
\begin{enumerate}
\item
\label{prop1}
the argument of the infrared logarithms factorized in the QCD form
factor $\Sigma[u;\,\alpha]$, 
\beq
L \, \equiv \, \log \frac{1}{u} \, \cong \, \log  \frac{Q^2}{m_X^2};
\eeq 
\item
\label{prop2}
the argument of the running coupling $\alpha \, = \, \alpha(Q)$, from which 
the form factor $\Sigma[u;\,\alpha(Q)]$ depends (as well as the 
coefficient function and the remainder function).
\end{enumerate}
An explicit check of property \ref{prop1}. has been obtained
verifying the consistency between the resummed expression 
on the r.h.s. of eq.~(\ref{tripla}) expanded up to $O(\alpha)$ 
and the triple distribution computed to the same order in \cite{ndf}.
Since the dependence of the coupling on the scale is a 
second-order effect, point \ref {prop2}. cannot be verified
with the above computation.
The possibility for instance that the hard scale $Q=w\, m_b$ 
in (\ref{tripla}) is fixed instead by the beauty mass $m_b$,
\begin{equation}
\label{wrong}
Q \, = \, m_b~~(???),
\end{equation}
cannot be explicitly ruled out by comparing with the $O(\alpha)$ triple
distribution, because
\begin{equation}
\alpha(w \, m_b) \, = \, \alpha(m_b) \, + \, O\left(\alpha^2\right).
\end{equation}
A second-order computation of the triple distribution is not available
at present and therefore a direct check of 2. is not possible.
An indirect check can however be obtained as explained in the following.
Any semileptonic spectrum can be obtained from the triple distribution 
in (\ref{tripla}) by phase-space integration.
In general, different results are obtained for a spectrum if the coupling is 
evaluated at the scale (\ref{adnauseam}) or instead for example at the scale 
(\ref{wrong}).
It is therefore sufficient to compute a spectrum in a single variable with
our resummation scheme and to compare with an explicit second-order calculation.
As far as we known, the only available second-order computation is that of the
$O(\alpha^2 \, n_f)$ corrections to the distribution in the light-cone
momentum $p_+$ \cite{phatspec}.
To compute the fermonic corrections, one simply inserts in the gluon lines of the 
first-order diagrams a fermion bubble, avoiding therefore the computation
of the complicated two-loop topologies.
It is however not straightforward to extract information about resummation
from a fermionic computation, as the latter is not consistent 
``at face value'' with the exponentiation property of the form factors
\footnote{Exponentiation in $QED$ originates from the independence of mutiple
soft-photon emissions. In $QCD$, multiple gluon emissions are 
instad not independent because of non-zero gluon color charge, but
exponentiation still holds because of the cancellation of the 
non-abelian correlation effects related to the inclusive character
of the form factors.}.
 
In sec.~\ref{secphat} we resum the spectrum in $p_+$ in next-to-leading
order ($NLO$), which involves an integration over the hard scale $Q$. 
We also give the coefficients of the large logarithms $\alpha^n \, \log^k m_b/p_+$ 
up to third order included. The only coefficient that we are
not able to compute is that of the single logarithm in third
order, $\alpha^3 \, \log m_b/p_+$, for which a three-loop computation is 
required.
These logarithmic terms can be used in phenomenological analyses 
which do not implement the full resummed formulas or can be compared 
with fixed-order computations as soon as the latter become available.
By comparing the fermionic contributions to the coefficients of the
infrared logarithms with the result of the Feynman diagram
computation, we obtain an explicit and non-trivial check of our
resummation scheme.

In sec.~\ref{secel} we resum to $NLO$ the charged lepton energy 
spectrum. 
Also in this case there is an integration over the hard scale
$Q$ and many considerations made for the $p_+$ distributions
can be repeated for this case as well.
The electron spectrum is one of the first quantities studied in 
semileptonic decays both theoretically and experimentally 
\cite{altetal, akhouri,babarel}, because it allows for a
determination of the Cabibbo-Kobayashi-Maskawa ($CKM$) matrix 
element $|V_{ub}|$.
To avoid the large backgroud from the decay
\begin{equation}
B \, \rightarrow \, X_c \, + \, l \, + \, \nu,
\end{equation}
it is usually necessary to restrict the analysis to electron energies above the 
threshold of the previous process --- recently somehow below --- 
i.e. to lepton energies in the interval
\begin{equation}
\frac{m_b}{2} \, \left( 1 \, - \, \frac{m_c^2}{m_b^2} \right)
\, \le \, E_e \, \le \, \frac{m_b}{2}.
\end{equation}
The actual hadron kinematics is obtained with the
replacements $m_c \to m_D$ and $m_b \to m_B$.
That gives a rather tight window in the end-point region,
\beq
\Delta E_e \, = \, \frac{m_c^2}{2m_b} \, \to \, 
 \frac{m_D^2}{2m_B} \, = \, 330~{\rm MeV},
\eeq
which is dominated by large logarithms and Fermi-motion effects.
We will show however that these long-distance phenomena are
expected to have a smaller effect on this distribution than
for example on the hadron mass spectrum.

In secs.~\ref{secphat} and \ref{secel} we also discuss the validity of the so-called 
Brodsky-Lepage-Mackenzie ($BLM$) scheme \cite{BLM} for the evaluation of spectra in 
the threshold region --- see also \cite{O7twoloops} for a discussion on the radiative case 
(\ref{startrd}).
In general, this scheme aims at an estimate of the full 
second-order corrections $O(\alpha^2)$ by means of the evaluation of the fermionic 
corrections, i.e. of the $O(\alpha^2 n_f)$ terms only.
The idea is that the fermionic corrections carry with them a factor $\beta_0$,
so after the fermionic computation, one makes the replacement:
\begin{equation}
n_f \, \rightarrow \, - \, \frac{3}{2}\left( 4 \pi \beta_0 \right)
\, = \, n_f \, - \, \frac{33}{2},
\end{equation}
where $4 \pi \beta_0 \, = \, \left(11/3 \, C_A \, - \, 4/3 \, T_R \, n_f \right)$ 
is the first coefficient of the $\beta$-function. 
$C_A \, = \, N_C \, = \, 3$ is the Casimir of the adjoint representation
of the color group $SU(3)_C$ and $T_R \, = \, 1/2$ is the trace 
normalization of the fundamental generators.
The $BLM$ approximation is often a rather good one in estimating 
total cross section and inclusive decay widths.
The total semileptonic decay width for example is known at present
to full order $\alpha^2$:
\beq
\frac{\Gamma}{\Gamma_0} 
\, = \, K(\alpha) 
\, = \, 1 \, + \, K^{(1)} \, \alpha \, + 
\, \left( n_f \, \hat{K}^{(2)} \, + \, \delta K^{(2)} \right) \, \alpha^2.
\eeq
The explicit value of $K^{(1)}$ has been given in eq.~(\ref{K1})
and the values of $\hat{K}^{(2)}$ and $\delta K^{(2)}$ can be found
in  \cite{timo}.
With the $BLM$ ansatz, $K^{(2)}$ is over-estimated  
by $\approx 15 \%$ with respect to the exact 
result (with $n_f \, = \, 4$), i.e. this approximation works pretty 
well in this case.

In sec.~\ref{nuovasec} we study the $BLM$ ansatz for the 
radiative and semileptonic distributions in the hadron invariant mass 
and for the semileptonic distribution in the variable $u$ defined above.
The conclusion is that The $BLM$ ansatz works in general rather well,
typically within an accuracy of $25\%$.

Finally, in sec.~\ref{conclude} we draw our conclusions and
we discuss the forthcoming application of our formalism to
phenomenology. 

\section{$\hat{p}_+$ distribution}
\label{secphat}

Recently, the spectrum has been computed in the normalized light-cone
momentum $\hat{p}_+$ \cite{phatspec} defined as
\begin{equation}
\label{phatdef}
\hat{p}_+ \,=\, \frac{E_X - |\vec{p}_X|}{m_b} \,=\, \frac{u\, w}{1+u} \, \simeq \, u\, w
~~~~~~~~~~~~~~~~~(0 \le \hat{p}_+ \le 1),
\end{equation}
where in the last member we have kept the leading term in the threshold region 
$u\ll 1$ only.
Even though this variable is not special in the present framework, 
let us discuss its resummation for the reasons discussed in the
introduction.
For notational simplicity let us make everywhere in this section the replacement
$\hat{p}_+ \, \rightarrow \, p$.
The derivation is similar to that of the resummed hadron mass distribution made
in \cite{noi2}.
The distribution in $p$ is obtained integrating the distribution in the 
hadronic variables $u$ and $w$:
\begin{equation}
\frac{1}{\Gamma} \, \frac{d\Gamma}{d\hat{p}_+} \, = 
\, \int_0^2 dw \, \int_{\max(0,w-1)}^1 \, du \,
\frac{1}{\Gamma} \,  \frac{d^2\Gamma}{du\,dw}
\, \delta\left( p - \frac{u\, w}{1+u} \right).
\end{equation}
For techinical reasons, it is actually more convenient to compute the partially-integrated 
distribution given by:
\begin{equation}
\label{evfracP}
R_P(p) \, \equiv \, \frac{1}{\Gamma} \, \int_0^p
\frac{d\Gamma}{dp'} \, d p'
\, = \,
\int du\, dw \, \frac{1}{\Gamma} \, 
\frac{d^2\Gamma}{du\,dw} \, \theta\left( p - \frac{u\, w}{1+u} \right).
\end{equation}
Let us replace the resummed expression for the distribution in the hadronic variables
$u$ and $w$ \cite{noi} on the r.h.s. of eq.~(\ref{evfracP}):
\begin{eqnarray}
\label{fineono}
R_P(p) &=&
\int_0^2 dw \, \int_{\max(0,w-1)}^1 \, du \, 
C_H(w;\,\alpha)\, \sigma\left[u,\,\alpha(w\,m_b)\right] \, \theta\left(p - \frac{u\, w}{1+u} \right)
\, + \, \nonumber\\
&+& \int_0^2 dw \, \int_{\max(0,w-1)}^1 \, du \, d_H(u,w;\,\alpha)
\, \theta\left( p - \frac{u\, w}{1+u} \right),
\end{eqnarray}
where:
\begin{itemize}
\item 
$C_H(w;\,\alpha)$ is a short-distance coefficient function having
the following expansion in powers of $\alpha$:
\begin{equation}
\label{CHw}
C_H(w;\,\alpha) \, = \, C_H^{(0)}(w) \, + \, \alpha \, C_H^{(1)}(w)
\, + \, \alpha^2 \, C_H^{(2)}(w) \, + \, O(\alpha^3),
\end{equation}
with
\begin{eqnarray}
\label{CHw1}
C_H^{(0)}(w) &=& 2w^2(3-2w);
\\
\label{CHw2}
C_H^{(1)}(w) &=& \frac{C_F}{\pi} \, 2w^2(3-2w)
\left[ {\rm Li}_2(w) \, + \,\log w\log(1-w) \, - \, \frac{35}{8} 
\, - \, \frac{9-4w}{6-4w} \log w \right].
\end{eqnarray}
${\rm Li}_2(w) =\sum_{n=1}^{\infty} w^n/n^2$ is the standard dilogarithm;
\item
$\sigma\left[u;\,\alpha\right]$ is the differential QCD form factor:
\begin{equation}
\label{zero}
\sigma\left[u,\,\alpha\right] \,  = \, \frac{d}{du} \, \Sigma[u;\,\alpha] 
\, = \, \delta(u) \, + \, O(\alpha);
\end{equation}
\item
$d_H(u,w;\,\alpha)$ is a short-distance remainder function whose
expansion starts at $O(\alpha)$ and which has an integrable singularity
for $u\rightarrow 0$.
The first-order correction for example has a logarithmic
singularity:\footnote{The explicit expression can be found in \cite{noi}.}
\begin{equation}
d_H^{(1)}(u,w) \, \approx \, \log u~~~~~~~~~~~~~~~~~{\rm for}~u \rightarrow 0.
\end{equation}
\end{itemize}
We are interested in the region 
\begin{equation}
p \, \ll \, 1.
\end{equation}
The integral of the remainder function on the r.h.s. of eq.~(\ref{fineono})
vanishes in the limit $p\rightarrow 0$, because:
\begin{equation}
\lim_{p\rightarrow 0} \, \int dw \, du \, d_H(u,w;\,\alpha)
\, \theta\left( p - \frac{u\, w}{1+u} \right)
\, = \, \int dw \, du \, d_H(u,w;\,\alpha)
\, \lim_{p\rightarrow 0} \, \theta\left( p - \frac{u\, w}{1+u} \right)
\, = \, 0.
\end{equation}
The limit $p\rightarrow 0$ can be taken inside the integral because the integrand 
has, as already said, at most an integrable singularity for $u\rightarrow 0$.
The first integral on the r.h.s. of eq.~(\ref{fineono}) can be decomposed as:
\begin{equation}
\int_0^1 dw \int_0^1 du
C_H(w;\alpha) \sigma\left[u,\alpha(w\,m_b)\right] \theta\left(p - \frac{u w}{1+u} \right)
+ \int_0^1 du \int_1^{1+u} dw 
C_H(w; \alpha) \sigma\left[u, \alpha(w\,m_b)\right] \theta\left(p - \frac{u w}{1+u} \right).
\end{equation}
The second integral vanishes in the limit $p\rightarrow 0$ because it
extends to an infinitesimal domain in $w$. The lowest-order term,
obtained replacing for $\sigma[u;\,\alpha]$ the last member of eq.~(\ref{zero}), 
identically vanishes.

To summarize, as far as logarithmically enhanced terms and constants terms
in the limit $p \rightarrow 0$ are concerned, we can make the following 
approximations:\footnote{It is remarkable that the variable $p$ keeps
unitary range after the simultaneous approximations 2. and 3.}
\begin{enumerate}
\item
neglect the remainder function $d_H(u,w;\,\alpha)$, whose contribution
will be includere later on together with contributions of similar size;
\item
approximate the integration domain with a unit square:
\begin{equation}
0 \, \le \, u, \, w \le 1; 
\end{equation}
\item
simplify the kinematical constraint according to the last member in
eq.~(\ref{phatdef}), since large logarithms of $p$ can only come 
from the region $u \ll 1$.
\end{enumerate}
We then obtain:
\begin{equation}
\label{newschemeidea}
R_P(p;\,\alpha) \, = \,  \int_0^p \, dw \, C_H\left(w;\,\alpha\right) \, 
\, + \, 
\int_p^1 \, dw \, C_H\left(w;\,\alpha\right) 
\Sigma\left[ p /w; \, \alpha(w \, m_b)\right] \, + \, O(p;\,\alpha),
\end{equation}
where $O(p;\,\alpha)$ denote non-logarithmic, small terms to be 
included later on by matching with the fixed-order distribution.
To the same approximation, i.e. up to terms which $O(p;\,\alpha)$, 
we can further simplify the distribution by neglecting the first
integral and integrating the second integrand down to $w=0$:
\begin{equation}
R_P(p;\,\alpha) \, = \,
\int_0^1 \, dw \, C_H\left(w;\,\alpha\right) \,
\Sigma\left[ p /w; \, \alpha(w \, m_b)\right] \, + \, O(p;\,\alpha),
\end{equation}
where $\alpha=\alpha(m_b)$.
The neglected term is indeed:
\begin{equation}
\int_0^p \, dw \, C_H\left(w;\,\alpha\right) \,
\Big\{ 1 \, - \, \Sigma\left[ p /w; \, \alpha(w \, m_b)\right] \Big\}.
\end{equation}
Expanding the form factor in powers of $\alpha(m_b)$, the above
expression reduces to a linear combination of integrals of the form
\begin{equation}
~~~~~~~~~~~~~~~~~~~~~~\int_0^p \, dw \, C_H\left(w;\,\alpha\right) \,\alpha^n \log^k (p) \, \log^l (w)
\, \rightarrow \, 0~~~~~~{\rm for}~p\rightarrow 0
~~~~~~~~~~~
(n \, \ge \, 1;\, k,\, l \, \ge \, 0).
\end{equation}
We have therefore shown that the neglected term is $O(p;\,\alpha)$.

\noindent
By insering the first-order expressions for the coefficient function $C_H(w;\,\alpha)$
and the universal QCD form factor $\Sigma(u;\,\alpha)$, we obtain for the event fraction 
$R_P(p;\,\alpha)$, apart from small terms $O(p;\,\alpha)$:
\begin{equation}
\int_0^1 \, dw \, C_H\left(w;\,\alpha\right) \,
\Sigma\left[ p /w ;\, \alpha(w \, m_b)\right] 
\, = \, 
1 \, - \, 
\frac{C_F\,\alpha}{\pi}
\left( 
\frac{1}{2} \, \log^2 p
\, + \, \frac{13}{6} \, \log p
\, + \, \frac{463}{144} 
\right)
\, + \, O(\alpha^2),
\end{equation}
where $C_F \, = \, \left(N_C^2-1\right)/(2N_C)$ is the Casimir of the fundamental
representation of the color group $SU(3)_c$ and $N_C \, = \, 3$ is the number of
colors.
Let us note that we have a different coefficient for the single
logarithm with respect to those ones in the
hadron mass distribution in the semileptonic decay (\ref{startsl}) 
or the radiative decay (\ref{startrd}).

\subsection{Minimal scheme}

We introduce a resummed form of the event fraction as:
\begin{equation}
\label{R_P}
R_P(p;\,\alpha) \,=\, C_P(\alpha) \, \Sigma_P(p;\,\alpha) \, 
+ \, D_P(p;\,\alpha).
\end{equation}
All the functions above have an expansion in powers of $\alpha$:
\begin{eqnarray}
C_P(\alpha) &=& 1 \, + \, \alpha \, C_P^{(1)} \, + \, \alpha^2 \, C_P^{(2)} 
\,+ \, O(\alpha^3);
\\
\Sigma_P(u;\,\alpha) &=& 1 \, + \, \alpha \, \Sigma_P^{(1)}(p) \, + \alpha^2 \, \Sigma_P^{(2)}(p) 
\, + \, O(\alpha^3);
\\
D_P(p;\,\alpha) &=& \alpha \, D_P^{(1)}(p) \, + \alpha^2 \, D_P^{(2)}(p) 
\, + \, O(\alpha^3).
\end{eqnarray}
Let us begin considering a minimal scheme, where only logarithms 
in $p$ are included in the effective form factor $\Sigma_P$.
Since we will consider a different scheme later on, let
us denote the quantities in the minimal scheme with a bar.
The first-order corrections to the form factor and the coefficient function
in the minimal scheme read:
\begin{eqnarray}
\bar{\Sigma}_P^{(1)}(p) &=&
- \, \frac{C_F}{\pi}
\left(
\, \frac{1}{2}\log^2 p
\, + \, \frac{13}{6}\log p
\right);
\nonumber\\
\bar{C}_P^{(1)} &=& - \,\frac{C_F}{\pi} \, \frac{463}{144} \, = - \, 1.36461.
\end{eqnarray}
Note that the correction to the coefficient function is rather large:
for $\alpha(m_b)=0.22$ it amounts to $\approx \, - \, 30\%$.

We now expand the resummed result in powers of $\alpha$ and
compare with the fixed order result, which is known to full order 
$\alpha$ \cite{phatspec}:
\begin{equation}
R_P(p;\alpha)\,=\, 1 \,+\, \alpha \, R_P^{(1)}(p) \, + \, \alpha^2 \, R_P^{(2)}(p) 
\, + \, O(\alpha^3),
\end{equation}
with
\begin{eqnarray}
R_P^{(1)}(p) &=&
- \, \frac{C_F}{\pi}
\bigg[
\, \frac{1}{2}\, \log^2 p
\, + \, \frac{13}{6}\,\log p 
\, + \, \frac{1}{2}\, \log^2 (p) \,p^3 \, (2 - p) 
\, + \, \frac{1}{12}\,\log (p) \, p \, (4 - 23\, p - 14\, p^2 + 7\, p^3) 
+\nonumber\\
&&~~~~~~ 
+ \, \frac{1}{144}  (1 - p)^2\, (463 + 246\, p - 240\, p^2 + 
              64\, p^3 - 7\, p^4 + 2\, p^5)
\bigg].
\end{eqnarray}
We obtain for the leading-order remainder function:\footnote{
To our knowledge, this is the only first-order remainder function
possessing a double logarithmic term (multiplied by positive powers
of the variable $p$).}
\begin{eqnarray}
\bar{D}_P^{(1)}(p) & = & 
- \, \frac{C_F}{\pi}
\bigg[
\, \frac{1}{2}\, \log^2 (p) \,p^3 \, (2 - p) 
\, + \, \frac{1}{12}\,\log (p) \, p \, (4 - 23\, p - 14\, p^2 + 7\, p^3) 
+\nonumber\\
&&~~~~~~ + \,
\frac{1}{144}  (1 - p)^2 \, p \, ( 246 - 240\, p + 64\, p^2 - 7\, p^3 + 2\, p^4)
\, - \, \frac{463}{144} p \left( 2 - p \right)
\bigg].
\end{eqnarray}
Since 
\begin{equation}
\label{normSigbar}
\bar{\Sigma}_P(1;\,\alpha)\, = \, 1, 
\end{equation}
taking $p=1$ in eq.~(\ref{R_P}), we obtain the following 
relation between the coefficient function 
and the remainder function in the upper endpoint:
\begin{equation}
\bar{C}_P(\alpha) \,  = \, 1 \, - \, \bar{D}_P(1;\,\alpha). 
\end{equation}
It is a trivial matter to verify that the above relation holds true for our first-order
expressions.

A definition of the minimal scheme which does not make reference
to the explicit perturbative expansion of $C_H$ and $\Sigma$
can be given as follows. 
Neglecting $O(p;\,\alpha)$ terms, we have that
\begin{equation}
\int_0^1 
dw \, C_H(w;\alpha) \, \Sigma\left[p/w;\,\alpha(w\,m_b)\right] \, = \, 
\bar{C}_P(\alpha) \, \bar{\Sigma}_P\left(p;\,\alpha\right), 
\end{equation}
so that the coefficient function in the minimal scheme can be defined taking
$p=1$ in the above equation and using eq.~(\ref{normSigbar}):
\begin{equation}
\bar{C}_P(\alpha) \, \equiv \, 
\int_0^1 dw \, C_H(w;\alpha) \, \Sigma\left[1/w;\,\alpha(w\,m_b)\right]. 
\end{equation}
The effective form factor then reads:
\begin{equation}
\label{defSigmaP}
\bar{\Sigma}_P\left(p;\,\alpha\right) \, \equiv \,
\frac{ \int_0^1 dw \, C_H(w;\alpha) \, \Sigma\left[p/w;\,\alpha(w\,m_b)\right] }
{\int_0^1 dw \, C_H(w;\alpha) \, \Sigma\left[1/w;\,\alpha(w\,m_b)\right]}.  
\end{equation}
Let us comment upon the above result:
\begin{itemize}
\item
The universal form factor $\Sigma(u;\,\alpha)$ is evaluated for an
unphysical arguement $u>1$ in eq.~(\ref{defSigmaP}), but this does
not cause any problem because the perturbative form factor is an
analytic function of $u$ which can be continued in principle
to any (complex) value of $u$;
\footnote{
the form factor $\Sigma(u;\,\alpha)$ has been given to $NNLO$ in eq.(125) 
of \cite{noi}, where it is written as a function of
\beq
\tau \, = \, \beta_0 \, \alpha(Q) \, \log \frac{1}{u}.
\eeq
For $u>1$ we have that $\tau<0$, but that does not produce any 
singularity in $\Sigma(u;\,\alpha)$.}

\item
The effective form factor $\bar{\Sigma}_P(p;\alpha)$ defined in eq.~(\ref{defSigmaP}) 
has a similar form to the effective form factor for the $t$ distribution 
$\bar{\Sigma}_T(t;\alpha)$ considered in \cite{noi2}, where $t=m_X^2/m_b^2$: 
the only differences are that in the numerator of the definition of $\bar{\Sigma}_T$,
$t/w^2$ is replaced by $p/w$ and in the denominator $1/w^2$ is replaced 
by $1/w$. The same considerations made for the $t$-distribution therefore
can be repeated for the $p$-distribution.
There are long-distance effects which cannot be extracted from the 
radiative decay (\ref{startrd}), related to the integration over the 
hadron energy $w$ in the parametric region
\begin{equation}
p \, \ll \, w \, \ll \, 1.
\end{equation}
The effects of this region are however suppressed by the coefficient
function:
\begin{equation}
C_H(w) \, \approx \, w^2 \, = \, \frac{Q^2}{m_b^2} \, \ll \, 1.
\end{equation}

\end{itemize}

\subsection{Higher orders}

Let us now discuss the higher orders.
In order to remove the exponential (trivial) iterations, let us define
the exponent of the effective form factor by means of the relation:
\begin{equation}
\label{valeancora}
\bar{\Sigma}_P \, = \, e^{ \bar{G}_P },
\end{equation}
with
\begin{equation}
\bar{G}_P\left(p;\,\alpha\right) \,=\, \sum_{n=1}^{\infty} \sum_{k=1}^{n+1} 
\bar{G}_{P n k}\, \alpha^n \, L_p^k
\end{equation}
where we have defined:
\begin{equation}
L_p \, \equiv \, \log \frac{1}{p} \, \ge \, 0.
\end{equation}
We find:
\begin{eqnarray}
\bar{G}_{P 12} &=& G_{12};
\\
\bar{G}_{P 11} &=& G_{11} + \frac{5}{12}\,A_1;
\\
\bar{G}_{P 23} &=& G_{23};
\\
\bar{G}_{P 22} &=& G_{22}+
\frac{7}{96}\,A_1^2 + 
  \frac{5}{24}\,A_1\,\beta_0;
\\
\bar{G}_{P 21} &=& G_{21} +
\frac{5}{12}\,A_2 - 
   \frac{5}{12}\,B_1\, \beta_0
   + A_1^2\,
   \left( - \frac{47}
        {432}  + 
     \frac{5}{12}\,z(2) \right) \, +  
\nonumber\\
&+& \frac{23}{144}  \,A_1\,\beta_0
+ \frac{7}{48} A_1\left(B_1 + D_1\right) 
+  A_1 \frac{C_F}{\pi}\,
        \left( \frac{547}{216} - 
          2\,z(3) \right);
\\
\bar{G}_{P 34} &=& G_{34};
\\
\bar{G}_{P 33} &=& G_{33} 
+ \frac{83}{5184} \, A_1^3
+ \frac{7}{96} \,A_1^2\,\beta_0
+ \frac{5}{36}\, A_1 \,\beta_0^2;
\\
\bar{G}_{P 32} &=& G_{32} +
\frac{5}{12} \,A_2\,\beta_0 -   
   \frac{5}{12}\,B_1\,\beta_0^2 \, +
  {A_1}^3\,
   \left( - \frac{1117}
        {20736}  + 
     \frac{7}{48}\,z(2) - 
     \frac{5}{12}\,z(3) \right) \, + 
\frac{7}{48} \, A_1 \, A_2 \, +
\nonumber\\ 
&+& \, A_1 \, 
\left( 
\frac{23}{144}\,\beta_0^2 + \frac{5}{24}\,\beta_1  
\right)        
-\frac{7}{96}\,A_1\,\beta_0\,\left(B_1-D_1\right) 
+  \, A_1 \, \beta_0 \, \frac{C_F}{\pi}\,
           \left( \frac{547}{432} - z(3) \right) \, +
\nonumber\\
&+& \frac{83}{1728} {A_1}^2\, \left(B_1+D_1\right)
+ {A_1}^2 \, \beta_0 \,
      \left( \frac{47}{864}  + 
        \frac{25}{24}\,z(2) \right) 
+ {A_1}^2 \, \frac{C_F}{\pi }\,
        \left( \frac{22747}{6912} - 3 z(4) \right).
\end{eqnarray}
The $G_{ij}$'s are the coefficients of the threshold logarithms 
in the exponent of the form factor $G$ for the radiative decay
(\ref{startrd}). They have been explicitly given,
together with all the constants $A_i$, $B_i$, $D_i$ and $\beta_i$ 
entering the above equations, 
in \cite{noi}\footnote{The next-to-next-to-leading-order (NNLO) coefficients $A_3$, $B_2$ 
and $D_2$, which are a necessary input for the determination of the $G_{ij}$'s, 
have been computed in the past few years in refs. \cite{A3}, \cite{B2} 
and \cite{gardi0} respectively.}.
$z(s) \, = \, \sum_{n=1}^{\infty} 1/n^s$ is the Riemann Zeta-function with
$z(2)\, = \, \pi^2/6 \, = \, 1.64493\cdots$, $z(3) \, = \, 1.20206\cdots$ and 
$z(4) \, = \, \pi^4/90 \, = \, 1.08232\cdots$.

Replacing explicit values for the coefficients $\bar{G}_{P ij}\ne G_{ij}$, 
we obtain:
\begin{eqnarray}
\bar{G}_{ P 11 } &=&  \frac{C_F}{\pi} \, \frac{13}{6};
\\
\bar{G}_{ P 22 } &=& \frac{C_F}{2\pi^2}
\left[
  C_A \left( \frac{25}{24} + \frac{z(2)}{2} \right)
+ C_F \left( \frac{7}{48} - z(2) \right)
- \frac{n_f}{4}  
\right];
\\
\bar{G}_{ P 21 } &=& \frac{C_F}{2\pi^2}
\Bigg[
  C_A \bigg( \frac{1253}{144} - \frac{13}{4} z(2) - \frac{z(3)}{2} \bigg)
+ C_F \bigg( \frac{1303}{288} + \frac{17}{6} z(2) - 3 z(3) \bigg)
%+ \nonumber\\
%& & ~~~~~ 
- n_f \left( \frac{113}{72} - \frac{z(2)}{3} \right)
\Bigg];
\\
\bar{G}_{ P 33 } &=& 
\frac{C_F}{\pi^3}
\, \Bigg[ \frac{n_f^2}{648} +
    C_F\,C_A\left(\frac{77}{1152} - \frac{11}{8} \, z(2) \right) + 
        C_F\,n_f\left(\frac{29}{576} + \frac{z(2)}{4}\right) +\nonumber\\ 
&+&     C_A\,n_f \left(\frac{185}{1296} - \frac{z(2)}{12}\right) + 
   C_A^2\left(-\frac{1589}{2592} + \frac{11}{24}\,z(2) \right) + 
    C_F^2\left(\frac{83}{5184} + \frac{z(3)}{3} \right)
  \Bigg];
\\
\bar{G}_{P 32} &=&
\frac{C_F}{\pi^3}
\, \Bigg[ \left(  \frac{359}{2592} - \frac{z(2)}{36} \right)\,n_f^2 +
    C_F\,C_A\left(\frac{64331}{41472} + \frac{313}{96}\,z(2)
    - \frac{11}{3}\,z(3) + \frac{5}{4}\,z(4)\right) +\nonumber\\ 
&+&        C_F\,n_f\left(-\frac{3829}{20736} - \frac{2}{3}\,z(2)
     + \frac{5}{12}\,z(3)    \right) + 
        C_F^2 \left(\frac{65381}{20736} + \frac{7}{48}\,z(2)
      - \frac{13}{6}\,z(3) - \frac{11}{4}\,z(4)   \right) + \nonumber\\ 
&+&        C_A\,n_f \left(-\frac{949}{648} + \frac{53}{144}z(2)
    - \frac{z(3)}{24}  \right) + 
        C_A^2 \left(\frac{34907}{10368} - \frac{293}{288}\,z(2)
      + \frac{77}{48}\,z(3) - \frac{11}{16}\,z(4)    \right) 
  \Bigg].
\end{eqnarray}
Let us make a few remarks:
\begin{itemize}
\item
As in the case of the distributions in the variables $u$ and $t$ studied in 
\cite{noi2}, the leading logarithms $\alpha^n\,L^{n+1}$ have the same coefficients
as in the hadron mass spectrum in the radiative decay (\ref{startrd}):
differences only occur in $NLO$ and beyond;
\item
The differences between the coefficients $G_{Pij}$  and the coefficients $G_{ij}$ 
of the radiative spectrum are much smaller than for the $u$ or the $t$ distribution
resummed in \cite{noi2}.
That is an indication that the long-distance effects in the $p$ distribution and in 
the radiative decay (\ref{startrd}) are not very different \cite{phatspec}.
\end{itemize}

\subsection{Check of Resummation Formula}

In this section we check our results for the resummed $p$ 
spectrum against an explicit computation of the fermionic 
corrections, i.e. of the $O(\alpha^2 \, n_f)$
contributions \cite{phatspec}.
The computation with Feynman diagrams of the $O(\alpha^2 n_f)$ 
terms involves: 
\begin{enumerate}
\item 
\label{one}
neglecting double emissions off the primary charges, 
which would give contributions proportional
to $\alpha^2$ without any $n_f$ factor.
The expansion up to $O(\alpha^2)$ of the r.h.s. of eq.~(\ref{valeancora}) 
can  be made by expanding it in powers of $\bar{G}_P$,
\begin{equation}
\label{aggiunta}
\bar{\Sigma}_P \, = \, 1 \, + \, \bar{G}_P \, + \, 
\frac{1}{2} \, \bar{G}_P^{\,2} \, + \, \cdots
\end{equation}
and keeping terms up to second order in $\alpha$ in $\bar{G}_P$
and up to first order in $\bar{G}_P^2$.
There is not any multiple emission left in the fermonic computation,
because one neglects the terms 
$\bar{G}_{P12}^{\, 2}, \, \bar{G}_{P11}^{\, 2} \, = \, O(n_f^{\, 0})$ in 
eq.~(\ref{aggiunta}).
That means that relation (\ref{valeancora}) is truncated to its
(trivial) first order:
\beq
\label{interpret}
~~~~~~~~~~~~~~~~~~~~~~~~~~~~~~~~~~~~~~~~~~~~~~~~~~~~~~
\hat{\Sigma}_P \, \simeq \, 1 \, + \, \hat{G}_P
~~~~~~~~~~~~~~~~~~~~~
\left[O(\alpha^2 \, n_f) ~ {\rm computation}\right];
\eeq
\item
\label{two}
taking into account the secondary branching of a gluon
into a (real or virtual) quark-antiquark pair,
\beq
\label{fermionico}
g^* \, \to \, q \, + \, \bar{q},
\eeq
while neglecting the (non-abelian) gluon splitting,
\beq
g^* \, \to \, g \, + \, g,
\eeq
Only a {\it single} secondary branching of abelian kind (\ref{fermionico}) 
is therefore allowed. 
\end{enumerate}
According to eq.~(\ref{interpret}), the best interpretation
of a fermionic computation is that of relating this result
directly to $G_P$ and not to $\Sigma_P$.

\noindent
Because of \ref{two}., a non-trivial strucure of $G$ emerges:
\beq
\hat{G}_P \, = 
\, G_{P12} \, \alpha \, L_p^2 \, + \,  G_{P11} \, \alpha \, L_p
\, + \, \hat{G}_{P23} \, \alpha^2 \, L_p^3 
\, + \, \hat{G}_{P22} \, \alpha^2 \, L_p^2
\, + \, \hat{G}_{P21} \, \alpha^2 \, L_p.
\eeq
An $O(\alpha^2 n_f)$ computation therefore does not provide any check of
the exponentiation property, but it gives some information about 
the structure of $G$.
The exponentiation property of form factors has already 
been checked with many second-order and third-order computations
and it is not our concern.
We are only interested in the structure of $G$ which is, as we have already
said, sensitive to the specific integration over the hard scale.

Our second-order coefficients $\bar{G}_{P2i}$ have a contribution proportional to $n_f$ 
which is in agreement with the explicit computation:
\begin{eqnarray}
\bar{G}_{P23} &=& \frac{1}{9\,\pi^2} \, n_f \, + \, O\left(n_f^{\,0}\right);
\\
\bar{G}_{P22} &=& - \, \frac{1}{6\,\pi^2} \, n_f \, + \, O\left(n_f^{\,0}\right);
\\
\bar{G}_{P21} &=& \left( \frac{1}{27}\, - \, \frac{113}{108\,\pi^2} \right)\, n_f 
                  \, + \, O\left(n_f^{\,0}\right),
\end{eqnarray}
where the numerical values of $C_F \, = \, 4/3$ and $C_A \, = \, 3$ have been replaced.
If we assume instead that the hard scale $Q$ in the coupling 
$\alpha \, = \, \alpha(Q)$ 
entering the resummed formula (\ref{tripla}) is equal to the heavy flavor mass $m_b$ 
and not to (two-times) the hadron energy $2E_X$, 
we obtain different fermionic contributions to $\bar{G}_{P 22}$ and $\bar{G}_{P 21}$ which are
in disagreemnt with the fixed-order computation:
\begin{eqnarray}
\bar{G}_{P22}^{\,(Q=m_b)} &=& - \, \frac{7}{27\,\pi^2} \, n_f \, + \, O\left(n_f^{\,0}\right);
\\
\bar{G}_{P21}^{\,(Q=m_b)} &=& \left( \frac{1}{27}\, - \, \frac{47}{81\,\pi^2} \right)\, n_f 
                  \, + \, O\left(n_f^{\,0}\right).
\end{eqnarray}
This provides us with an explicit check of the resummation formula for the
triple-differential distribution at the two-loop level.
Let us remark that this check is not related to the study of the $BLM$
scheme: we just compared the coefficients of two independent $O(\alpha^2 n_f)$
computations and we never made the $BLM$ ansatz.
Our aim indeed was simply that of studying the effects of eq.~(\ref{adnauseam})
in a partial contribution, not that of estimating full second-order
corrections.

\subsection{The BLM scheme for the form factor}
\label{secBLM}

In this section we compare our exact results for the resummed spectrum in $p$
with those ones coming from the $BLM$ ansatz.
Let us make a comparison of the coefficients $G_{P2i}$ in $QCD$ in the
minimal scheme defined above with the
$BLM$ ones $\hat{G}_{P2i}$ starting from the leading terms :
\begin{enumerate}
\item
$\alpha^2 \, L_p^3$; the $QCD$ coefficient
\begin{equation}
\bar{G}_{P 23} \, = \,
- \, \frac{11}{6 \, \pi^2} \, + \, \frac{n_f}{9 \, \pi^2} 
\,  = \, - \, 0.185756 \, + \, 0.0112579 \, n_f
\end{equation}
is in complete agreement with the $BLM$ estimate:
\begin{equation}
\hat{G}_{P 23} \, = \, \bar{G}_{P 23}.
\end{equation}
The approximation turns out to be exact because the $QCD$ 
coefficient is proportional, as it is well known \cite{abcmv}, 
to the first coefficient of the $\beta$-function $\beta_0$;
\item
$\alpha^2 \, L_p^2$; the $QCD$ coefficient is
\begin{equation}
\bar{G}_{P 22} \, = \,
\frac{1}{54} \, + \, \frac{239}{108 \, \pi^2} \, - \, 
\frac{n_f}{6 \, \pi^2}
\, = \, 0.242739 \, - \, 0.0168869 \, n_f
\end{equation}
to be compared to the $BLM$ estimate:
\beq
\hat{G}_{P 22} \, = \,  -\,\frac{1}{6\pi}\,\left(n_f-\frac{33}{2}\right)
\, = \, 0.278633\, -\, 0.0168869 \,n_f.
\eeq
The $BLM$ ansatz over-estimates the non-abelian contribution
by $\approx \, 15 \%$, i.e. it works rather well;
\item
$\alpha^2 L_p$; in $QCD$ we have the value:
\begin{equation}
\bar{G}_{P 21} \, = \, - \, \frac{215}{324} \, + \, \frac{13883}{648 \, \pi^2} 
\, - \, \frac{11 \, z(3)}{3 \, \pi^2} + 
\left( \frac{1}{27} \, - \, \frac{113}{108 \, \pi^2} \right) \, n_f
\, = \, 1.06059 \, - \, 0.0689749 \, n_f 
\end{equation}
to be compared to the $BLM$ estimate
\beq
\hat{G}_{P 21} \, = \,   \left( \frac{1}{27} \, - \, \frac{113}{108 \, \pi^2}
\right) \,\left(n_f-\frac{33}{2}\right)
\, = \,1.13809\,- \,0.0689749\, n_f.
\eeq
The $BLM$ ansatz works with an accuracy better than $10\%$.
Let us note also that the $BLM$ cannot reproduce the terms proportional
to the trascendental constant $z(3)$ defined in the previous section.
\end{enumerate}
We conclude that the $BLM$ scheme works rather well for the $p$
spectrum.

\subsection{Non minimal scheme}

In this section we consider a non-minimal factorization scheme
which has various advantages.
The event fraction has a resummed form equal to the one
in the minimal scheme,
\begin{equation}
R_P(p;\,\alpha) \, = \, C_P(\alpha) \, \Sigma_P( p;\, \alpha )
\, + \, D_P( p;\, \alpha ),
\end{equation}
with different expressions for the coefficient function, the
form factor and the remainder function.
As in the minimal scheme, all these function have a perturbative expansion
in $\alpha$:
\begin{eqnarray}
C_P\left(\alpha\right) & = & 1 \, + \, \alpha \, C_P^{(1)}
\, + \, \alpha^2 \, C_P^{(2)} \, + \, O\left(\alpha^3\right);
\\
\Sigma_P\left(p;\,\alpha\right) & = & 1 \, + \, \alpha \, \Sigma_P^{(1)}(p)
\, + \, \alpha^2 \, \Sigma_P^{(2)}(p) \, + \, O\left(\alpha^3\right);
\\
D_P\left(p;\,\alpha\right) & = & \alpha \, D_P^{(1)}(p)
\, + \, \alpha^2 \, D_P^{(2)}(p) \, + \, O\left(\alpha^3\right).
\end{eqnarray}
The coefficient function is defined, to all orders, as:
\begin{equation}
C_P\left(\alpha\right) \, = \, \int_0^1 \, dw \, C_H\left(w;\,\alpha\right),
\end{equation}
and the expression for the effective form factor is:
\begin{equation}
\label{SigmaPeff2}
\Sigma_P\left(p;\,\alpha\right) \, = \,
\frac{ \int_0^1 \, dw \, C_H\left(w;\,\alpha\right) \,
\tilde{\Sigma}\left[ p /w ;\,\alpha(w \, m_b)\right] }
{ \int_0^1 \, dw \, C_H\left(w;\,\alpha\right) }.
\end{equation}
$\tilde{\Sigma}\left[u ;\,\alpha\right]$ is the extended form
factor defined in \cite{noi2}, equal to the standard one 
$\Sigma\left[u ;\,\alpha\right]$ for argument less than
one, $u<1$, and equal to one for a larger argument, $u\ge 1$.
Explicitly (cfr. eq.~(\ref{newschemeidea}))
\footnote{Replacing the extended form factor $\tilde{\Sigma}$ 
with the usual one $\Sigma$ in eq.~(\ref{SigmaPeff2}) amounts to 
dropping the infinitesimal terms for $p\rightarrow 0$ in $\Sigma_P$.}:
\begin{equation}
\Sigma_P\left(p;\,\alpha\right) \, = \,
\frac{ 
\int_0^p \, dw \, C_H\left(w;\,\alpha\right) \, + \, 
\int_p^1 \, dw \, C_H\left(w;\,\alpha\right) \,
\Sigma\left[ p /w ;\,\alpha(w \, m_b)\right] 
}{ \int_0^1 \, dw \, C_H\left(w;\,\alpha\right) }.
\end{equation}
This effective form factor is normalized, to all orders, as in the 
minimal scheme:
\begin{equation}
\Sigma_P\left(1;\,\alpha\right) \, = \,1.
\end{equation}
By inserting the expansion for $C_H(w;\,\alpha)$ and for 
$\tilde{\Sigma}(u;\,\alpha)$ and integrating over the hadronic energy $w$, 
we obtain for the first-order terms:
\begin{equation}
C_P^{(1)} \,=\, - \,\frac{C_F}{\pi} \, \frac{335}{144} \, = \, - \, 0.98735
\end{equation}
and
\begin{equation}
\Sigma_P^{(1)}(p) \, = \, - \,
\frac{C_F}{\pi}
\left(
\frac{1}{2} \log^2 p + \frac{13}{6} \log p
+ \frac{8}{9} - \frac{25}{18} p^3 + \frac{1}{2} \, p^4
\right).
\end{equation}
The coefficient function has a smaller value in this scheme
than in the minimal one and the effective form factor
contains not only a constant term for $p\rightarrow 0$
but also infinitesimal terms in the same limit.
Matching with the fixed-order distribution as we have done
in the minimal case, we obtain for the remainder function:
\begin{eqnarray}
D_P^{(1)}(p) &=& \frac{C_F}{\pi}
\Big[
\, - \, \frac{1}{2} (2 - p)\, p^3\, \log^2 p
\, - \, \frac{1}{12} (4 - 23\, p - 14\, p^2 + 7\, p^3)\,p\, \log p 
\, + \, 
\nonumber\\ 
&&~~~~~ 
+ \, \frac{p}{144} \left(
680 + 269\, p - 990\, p^2 + 447\,p^3 - 80\, p^4 + 11\, p^5 - 2\, p^6 
\right)
\Big].
\end{eqnarray}
Let us note that, with our scheme choices,
\begin{equation}
C_P(\alpha) \, = \, C_T(\alpha) = \, C_U(\alpha)
\end{equation}
to all orders in $\alpha$, where $C_T(\alpha)$ and $C_U(\alpha)$ are 
the coefficient functions for the distributions 
in the variables $u$ and $t$ computed in \cite{noi2}.

The exponent of the effective form factor in this non-minimal scheme
$G_P$ contains, in the limit $p\rightarrow 0$, also constants terms
($\bar{G}_{Pij}=G_{Pij}$ for $j\ge 1$):
\begin{eqnarray}
G_{P10} &=& - \, \frac{23}{144} \, A_1 \, + \, \frac{5}{12} \, \left(B_1 \, + \, D_1\right);
\\
G_{P20} &=& 
- \, \frac{23}{144}\,A_2 - 
  \frac{101}
   {576}\,A_1\,\beta_0 + \frac{23}{48}\,\beta_0\,
     \left( B_1 + 
       \frac{2}{3}\,D_1 \right)
   - \frac{47}{432}\,A_1\,
     \left( B_1 + D_1
       \right)  + 
  \frac{7}{96}\,{\left( B_1 + 
         D_1 \right) }^2 
\, + \nonumber\\
&+&
\frac{5}{12}\,\left( B_2 + 
       D_2 \right)   +  
  \frac{5}{12} \,z(2)\,A_1\,
     \left( B_1 + D_1
       \right) 
+ B_1\, \frac{C_F}{\pi }\,
     \left( \frac{547}{216} - 2\,z(3) \right)  
+ D_1\, \frac{C_F}{\pi } \,
     \left( \frac{547}{216} - 2\,z(3) \right)  
+ \nonumber\\
&+& {A_1}^2\,
   \left( \frac{2057}{41472} - 
     \frac{23}{144}\,z(2) + \frac{5}{12}\,z(3)
     \right)  \, + \, A_1 \, \frac{C_F}{\pi } \,
     \left( - \frac{90121}{20736}
           + \frac{5}{6}\,z(3) + 3\,z(4) \right).
\end{eqnarray}
Explicitly:
\begin{eqnarray}
G_{P10} &=& - \, \frac{8}{9} \, \frac{C_F}{\pi};
\\
G_{P20} &=& 
\frac{C_F}{{\pi }^2} \,
\Bigg[
    \, n_f \,
       \left( \frac{2243}{5184} - 
         \frac{5}{72}\,z(2) \right)  + 
      C_F\,
       \left( - \frac{14435}{1728}
             - \frac{83}{144}\,z(2) + 
         \frac{19}{4}\,z(3) + 3\,z(4) - 
         \frac{5}{8}\,z(3) \right)
\, + \nonumber\\          
&& ~~~~~ + \, C_A\,
       \left( - \frac{24775}{10368}
             + \frac{193}{288}\,z(2) + 
         \frac{5}{48}\,z(3)
         \right)
\Bigg].
\end{eqnarray}
As in the case of the $t$ distribution considered in \cite{noi2},
the coefficients functions are related in the two schemes
by the following relations:
\begin{eqnarray}
C_{P}^{(1)} &=& \bar{C}_{P}^{(1)} - G_{P10};
\\
C_{P}^{(2)} &=& \bar{C}_{P}^{(2)} - \bar{C}_{P}^{(1)} G_{P10}
+ \frac{1}{2} G_{P10}^2 - G_{P20}.
\end{eqnarray}
The first of the above equations is easily verified by inserting
our first-order expressions.

\section{Electron spectrum}
\label{secel}

In our framework, the electron spectrum is obtained by integrating the distribution 
in the hadron and electron energies $w$ and $\overline{x}\equiv 1-x$, resummed to NLO in \cite{noi}, 
over the hadron energy:
\begin{equation}
\frac{1}{\Gamma}\frac{d\Gamma}{d\bar{x}}
\, = \, 
\int_{\bar{x}}^{1 \, + \, \bar{x}} dw \,
\frac{1}{\Gamma} \, \frac{d^2\Gamma}{dw d\bar{x}}.
\end{equation}
In the previous case, in order to avoid generalized functions
entering the differential form factor $\sigma[u;\,\alpha]$,
we have considered the partially integrated distribution or event fraction.
That in not necessary in this case because the differential electron spectrum,
as we are going to show, already contains the partially integrated form factor 
$\Sigma[u;\,\alpha]$.
Inserting the resummed expression for the double distribution, 
we obtain:\footnote{The explcit expressions of the coefficients functions and the 
remainder functions can be found in \cite{noi}.}
\begin{eqnarray}
\label{almostex}
\frac{1}{\Gamma}\frac{d\Gamma}{d\bar{x}}
&=& 
\int_{\bar{x}}^1 dw \,C_L(\bar{x},w; \alpha)
\Sigma\left[ \frac{ \bar{x} }{  w } ; \alpha(w\,m_b) \right] +
\int^{1+\bar{x}}_1 dw
\,C_{XW1}(\bar{x};\alpha)
\Big\{
1  -  C_{XW2}(\bar{x};\alpha)  \Sigma\left[w-1; \alpha(m_b) \right]
\Big\}
\nonumber\\
&+& \int_{\bar{x}}^1 dw \, d_<(\bar{x},w;\,\alpha)
\, + \, \int^{1+\bar{x}}_1 dw \, d_>(\bar{x},w;\,\alpha).
\end{eqnarray}
We are interested in the region 
\begin{equation}
\label{smallxb}
\bar{x} \, \ll \, 1,
\end{equation} 
which selects hadron final states with a small invariant mass
and produces large logarithms in the perturbative expansion. 
The integral of the first remainder function $d_<(\bar{x},w;\,\alpha)$
vanishes in the limit $\bar{x}\rightarrow 0$:
\begin{equation}
\int_{\bar{x}}^1 dw \, d_<(\bar{x},w;\,\alpha) \, \simeq
\int_0^1 dw \, d_<(\bar{x},w;\,\alpha) \, \rightarrow \, 0 ~~~~~~{\rm for}~ 
\bar{x}\rightarrow 0,
\end{equation}
because:
\begin{equation}
~~~~~~~~~~~~~~~~~~~~~~~~~~~~~~~~~~~~~~~~~ d_<(\bar{x},w;\,\alpha) \, \rightarrow \, 0 
~~~~~~~~~{\rm for}~ 
\bar{x} \, \rightarrow \, 0 ~~~~~~~~~~~~~~~~~~~~~~~~~~~~~~~ (0 < w < 1).
\end{equation}
The integral of the second remainder function $d_>(\bar{x},w;\,\alpha)$
also vanishes in the same limit: 
\begin{equation}
\int^{1+\bar{x}}_1 dw \, d_>(\bar{x},w;\,\alpha)\, \rightarrow \, 0 ~~~~~~~{\rm for}~ 
\bar{x} \, \rightarrow \, 0,
\end{equation}
because the integral extends to an infinitesimal region and the integrand vanishes 
for $w \rightarrow 1^+$:
\begin{equation}
d_>(\bar{x},w;\,\alpha) \, \rightarrow \, 0 ~~~~~~{\rm for}~ 
w \, \rightarrow \, 1^+.
\end{equation}
Finally, the integral involving $\Sigma[w-1;\alpha]$ also vanishes
because, after expanding the form factor in powers of $\alpha$, it
produces terms of the form:
\begin{equation}
\int_1^{1+\bar{x}} \, dw \, \alpha^n \log^k(w-1)\, \rightarrow \, 0 ~~~~~~~~~{\rm for}~ 
\bar{x} \rightarrow 0,
\end{equation}
because the logarithm function to any positive power $k \ge 0$, $\log^k(u)$, has an integrable 
singularity in $u=0$. We may write therefore:
\begin{equation}
\label{ameta}
\frac{1}{\Gamma}\frac{d\Gamma}{d\bar{x}}
\, = \, 
\int_{\bar{x}}^1 dw \,C_L(\bar{x},w;\,\alpha)
\,\Sigma\left[ \frac{ \bar{x} }{  w } ; \,\alpha(w\,m_b) \right]
\, + \,O(\bar{x};\alpha),
\end{equation}
where by $O(\bar{x};\alpha)$ we denote terms which vanish for $\bar{x}\rightarrow 0$
as well as for $\alpha\rightarrow 0$ (the neglected terms also vanish
for $\alpha\rightarrow 0$).
Let us conclude that, as long as logarithmically enhanced terms for $\bar{x}\rightarrow 0$
and constants are concerned, the integration can be made in the leading region
\begin{equation}
\label{leadreg}
0 \, \le \, w \, \le \, 1,
\end{equation}
which coincides with the phase-space domain at the tree level.

\subsection{Minimal scheme}

For clarity's sake let us consider at first a minimal factorization
scheme.
Since we are in region (\ref{smallxb}), we can take the limit 
$\bar{x}\rightarrow 0$ in the r.h.s. of eq.~(\ref{ameta}) anytime 
we do not encounter singularities:
\begin{equation}
\frac{1}{2\Gamma}\frac{d\Gamma}{d\bar{x}}
\, = \, \frac{1}{2}
\int_0^1 dw \,C_L(0,w;\,\alpha)
\,\Sigma\left[ \frac{ \bar{x} }{  w } ; \,\alpha(w\,m_b) \right]
\, + \, O(\bar{x}),
\end{equation}
where we have divided by a factor two in order to simplify the forthcoming
formulas. Let us note that, since we have taken the limit $\bar{x}\rightarrow 0$
in the coefficient function $C_L(\bar{x},w;\,\alpha)$ which also has an $O(\alpha^0)$
contribution and we have integated down to $w=0$, 
we have modified the integral above by terms $O(\bar{x})$ not multiplied by $\alpha$:
we will take care of them in a moment.
Replacing the first-order expressions for the coefficient function \cite{noi2}, 
\begin{equation}
C_L(0,w;\,\alpha)
\, = \,
12 \, w \, (1\, - \, w)
\Bigg\{ 1 \, + \,
\frac{\alpha\,C_F}{\pi} 
\left[
{\rm Li}_2(w) + \log w \log(1-w)
-\frac{3}{2}\log w - \frac{w \log w}{2(1-w)} - \frac{35}{8}
\right]
\, + \, O\left(\alpha^2\right)
\Bigg\},
\end{equation}
and the QCD form factor given in eq.~(\ref{Sigma}), we obtain:
\begin{equation}
\frac{1}{2}
\int_0^1 dw \,C_L(0,w;\,\alpha)
\,\Sigma\left[ \frac{ \bar{x} }{  w } ; \,\alpha(w\,m_b) \right]
\, = \, 
1 \, - \, \frac{\alpha_S\,C_F}{\pi}
\left(
\frac{1}{2} \log^2\bar{x}
\, + \, \frac{31}{12} \, \log\bar{x}
\, + \, \frac{15}{4}
\right)
\, + \, O(\alpha^2).
\end{equation}
In order to avoid a remainder function $O(\alpha^0)$, let us introduce
an over-all factor equal to the lowest-order spectrum, so that the
resummed form for the electron spectrum reads:
\begin{equation}
\frac{1}{2\,\Gamma}\frac{d\Gamma}{d\bar{x}} \, = \, 
(1\, - \, \bar{x})^2 \, (1\, + \, 2 \, \bar{x})
 \, \Big[ \, \bar{C}_X(\alpha) \, \bar{\Sigma}_X(\bar{x};\,\alpha) 
\, + \, \bar{d}_X(\bar{x};\,\alpha) \, \Big].
\end{equation}
We require that the remainder function vanishes for $\bar{x}\rightarrow 0$:
\begin{equation}
\lim_{\bar{x}\rightarrow 0} \bar{d}_X(\bar{x};\,\alpha) \, = \, 0.
\end{equation}
The coefficient function,
\begin{equation}
\bar{C}_X(\alpha) \, = \, 1 \, + \,
\alpha \, \bar{C}_X^{(1)}
 \, + \, \alpha^2 \, \bar{C}_X^{(2)},
\, + \, O(\alpha^3)
\end{equation}
has the leading correction:
\begin{equation}
\bar{C}_X^{(1)} \, = \, - \frac{15\,C_F}{4\,\pi} \, = \, - \, \frac{5}{\pi} \, 
= \, - \, 1.59155. 
\end{equation}
Let us note that this correction is very large as it amounts to
$\approx - 35\%$ for $\alpha(m_b)=0.22$.
The effective electron form factor,
\begin{equation}
\bar{\Sigma}_X(\bar{x};\,\alpha)
\, = \,
1 \, + \, \alpha \, \bar{\Sigma}_X^{(1)}(\bar{x})
\, + \, \alpha^2 \, \bar{\Sigma}_X^{(2)}(\bar{x})
\, + \, O(\alpha^3),
\end{equation}
has a first-order expression:
\begin{equation}
\bar{\Sigma}_X^{(1)}(\bar{x}) \,  = \,
- \, \frac{C_F}{2\,\pi } \, \log^2 \bar{x}
\, - \, \frac{31\,C_F}{12\,\pi }\,\log\bar{x}. 
\end{equation}
Let us now compute the remainder function, which is necessary
to describe also the region $\bar{x}\sim O(1)$ and
to have a uniform approximation in the whole $\bar{x}$ domain.
The remainder function has, as usual, an expansion of the form:
\begin{equation}
\bar{d}_X(\bar{x};\,\alpha) \, = \,
\alpha \, \bar{d}_X^{(1)}(\bar{x})
\, + \, \alpha^2 \, \bar{d}_X^{(2)}(\bar{x})
\, + \, O(\alpha^3).
\end{equation}
We match with the fixed-order spectrum,
\begin{equation}
\frac{1}{2\,\Gamma} \, \frac{d\Gamma}{d\bar{x}}
\, = \, (1-\bar{x})^2 \, (1\, + \, 2 \, \bar{x}) \,
\left[
1 \, + \, \alpha_S \, H^{(1)}(\bar{x}) 
\, + \, \alpha_S^2 \, H^{(2)}(\bar{x})
\, + \, O(\alpha^3)
\right],
\end{equation}
whose first order correction is well known \cite{kuhn,ndf}:
\begin{eqnarray}
 H^{(1)}(\bar{x}) &=&
\frac{C_F}{\pi}
\Bigg\{
\, - \, \frac{1}{2} \, \log^2 \bar{x}
\, - \, \left[ \, 
     \frac{2}{3} \, + \, 
     \frac{41}{36\,(1-\bar{x})^2} \, - \,
     \frac{13}{54\,(1-\bar{x})} \, + \,
     \frac{55}{54\,\left( 1 + 2\,\bar{x} \right) } 
\, \right] 
\, \log \bar{x}
\, + \nonumber\\
&& ~~~~~~~ - \, \frac{4}{3} \, - \, 
  \frac{41}{36\,(1-\bar{x})} \, - \,  
  \frac{23}
   {18\,\left(  1 + 2\,\bar{x} \right) } 
\, + \, \log\bar{x} \, \log(1-\bar{x}) + {\rm Li}_2(\bar{x})
\Bigg\}.
\end{eqnarray}
The leading-order remainder function reads:
\begin{eqnarray}
\bar{d}_X^{(1)}(\bar{x}) &=& 
\frac{C_F}{\pi} \,
\Bigg\{
\, \log \bar{x} \, \log (1-\bar{x}) + {\rm Li}_2(\bar{x})
\, + \, \frac{29}{12} \, - \, \frac{41}{36(1-\bar{x})} 
\, - \, \frac{23}{18(1+2\bar{x})} \, +
\nonumber\\
&&~~~~ + \,\left[ \frac{23}
      {12} - \frac{41}
      {36\,(1-\bar{x})^2} + 
     \frac{13}{54\,(1-\bar{x})} - 
     \frac{55}
      {54\,\left( 1 + 2\,\bar{x}
          \right) } \right] \, \log \bar{x}
\,\Bigg\}.
\end{eqnarray}
It vanishes, as requested, for $\bar{x}\rightarrow 0$ and,
despite appearances, it has not any $1/(1-\bar{x})$ singularity.

\noindent
The coefficient function can be defined to all orders as:
\begin{equation}
\bar{C}_X(\alpha) \, = \, 
\frac{1}{2}
\int_0^1 dw \,C_L(0,w;\,\alpha)
\,\Sigma\left[ \frac{ 1 }{  w } ; \,\alpha(w\,m_b) \right]
\end{equation}
and the effective form factor as:
\begin{equation}
\label{richiamo}
\bar{\Sigma}_X(\bar{x};\,\alpha) \, = \, 
\frac{ \int_0^1 dw \,C_L(0,w;\,\alpha)
\,\Sigma\left[ \, \bar{x} / w ; \,\alpha(w\,m_b) \right] }
{ \int_0^1 dw \,C_L(0,w;\,\alpha)
\,\Sigma\left[ \, 1 / w ; \,\alpha(w\,m_b) \right] }.
\end{equation}
The latter is normalized to all orders as:
\begin{equation}
\bar{\Sigma}_X(1;\,\alpha) \, = \,1.
\end{equation}
The effective form factor (\ref{richiamo}) factorizes the long-distance
effects related to the threshold region, a part of which cannot be
extracted from the radiative decay (\ref{startrd}).
An integration over the hard scale $Q=w\,m_b$ is indeed involved
in $\bar{\Sigma}_X$ while, in the radiative decay, kinematics 
fixes $w=1$ and there is no such integration.
The relevant integration region is, parametrically:
\begin{equation}
\bar{x} \, \ll w \, \ll \, 1,
\end{equation}
where the large logarithms $|\log(\bar{x}/w)| \gg 1$ are multiplied by a coupling 
$\alpha(w\, m_b) \gg \alpha(m_b)$.
As in the case of the $p_+$ distribution, the effects of this
region are however suppressed by the coefficient function:
\begin{equation}
C_L(w) \, \approx \, w \, = \, \frac{Q}{m_b} \, \ll \, 1. 
\end{equation}

\subsection{Higher orders}

Let us write as usual
\begin{equation}
\bar{\Sigma}_X \, = \, e^{\bar{G}_X},
\end{equation}
where the exponent of the form factor has the perturbative expansion:
\begin{equation}
\bar{G}_X\left(\bar{x};\,\alpha\right) \,=\, \sum_{n=1}^{\infty} \sum_{k=1}^{n+1} 
\bar{G}_{X n k}\, \alpha^n \, L_{\bar{x}}^{\,k},
\end{equation}
with
\begin{equation}
L_{\bar{x}} \, \equiv \, \log \frac{1}{\bar{x}} \, \ge \, 0.
\end{equation}
By inserting the truncated expansions for the coefficient function $C_L$ and the form
factor $\Sigma$ in eq.~(\ref{richiamo}), expanding the product, integrating term by term
and taking the logarithm, we obtain:
\begin{eqnarray}
\bar{G}_{X 12}&=&G_{12};
\\
\bar{G}_{X 11}&=&G_{11}+\frac{5}{6} A_1;
\\
\bar{G}_{X 23}&=&G_{23};
\\
\bar{G}_{X 22}&=&G_{22}+\frac{13}{72} A_1^2 + \frac{5}{12} A_1 \beta_0;
\\
\bar{G}_{X 21}&=&G_{21}
+ \frac{5}{6} A_2 + A_1^2 \left[ \frac{5}{6} z(2) -\frac{25}{54} \right]
+ \frac{19}{36} A_1 \beta_0
+ \frac{13}{36} A_1 (B_1+D_1) +
\nonumber\\
&-& \frac{5}{6} \beta_0 \, B_1
+ A_1 \frac{C_F}{\pi} \left[\frac{583}{216} - 2 z(3)\right];
\\
\bar{G}_{X 34}&=& G_{34};
\\
\bar{G}_{X 33}&=& G_{33} + \frac{35}{648} A_1^3 + \frac{13}{72} A_1^2 \, \beta_0
+ \frac{5}{18} \, A_1 \, \beta_0^2;
\\
\bar{G}_{X 32}&=& G_{32} +
\frac{5}{6} \, A_2 \, \beta_0 - \frac{5}{6}\,{\beta_0}^2\,B_1 
+\frac{13}{36} \, A_1 \, A_2 
+ \, A_1 \, \left( \frac{19}{36} \, {\beta_0}^2 \, + 
     \frac{5}{12} \, \beta_1 \right) \, +
\nonumber\\
&-& \frac{13}{72} A_1 \, \beta_0\,
      \left( B_1 - D_1 \right) + 
        A_1 \, \beta_0\,\frac{C_F}{\pi }\,
           \left( \frac{583}{432} - 
             z(3) \right) +
{A_1}^2\,\frac{C_F }{\pi }\,
        \left( \frac{737}{216} - 
          3\,z(4) \right) \, +
\nonumber\\
&+& 
  {A_1}^3\, \left( -  \frac{5}{16} + 
     \frac{13}{36} \,z(2) - 
     \frac{5}{6} \, z(3) \right)  + 
  \frac{35}{216} {A_1}^2\,
   \left( B_1 + D_1 \right) + 
     {A_1}^2\, \beta_0\,
      \left( \frac{25}{108} + 
        \frac{25}{12} \, z(2) \right).
\end{eqnarray}
Explicitly, one has for the coefficients $\bar{G}_{X ij} \ne G_{ij}$:
\begin{eqnarray}
\bar{G}_{X 11}&=&\frac{31\,C_F}{12\,\pi };
\\
\bar{G}_{X 22}&=&
\frac{C_F }{\pi^2}
\, \left[  
- \frac{23}{144} \,n_f 
+  C_F \,\left(\frac{13}{72} - \frac{z(2)}{2}\right) 
+ C_A\,
       \left( \frac{205}{288} + 
         \frac{z(2)}{4} \right) 
      \right];
\\
\bar{G}_{X 21}&=&
\frac{C_F }{\pi^2}
\left[ n_f
       \left( - \frac{73}
            {72}  + 
         \frac{z(2)}{6} \right) 
       + C_F
       \left( \frac{163}{96} + 
         \frac{11}{6} z(2) - 
         \frac{3}{2} z(3) \right)  + 
      C_A
       \left( \frac{23}{4} - 
         \frac{11}{6} z(2) - 
         \frac{z(3)}{4}
         \right)  \right];
\\
\bar{G}_{X 33}&=&
\frac{C_F}{\pi^3}
 \Bigg[\frac{7}{1296}\,n_f^2 +
    C_F\,C_A\left(\frac{143}{864} - \frac{11}{8}\,z(2)\right) + 
       C_F\,n_f\left(\frac{7}{216} + \frac{z(2)}{4}\right) + 
          C_A\,n_f\left(\frac{65}{648} - \frac{z(2)}{12}\right)+\nonumber\\ 
&+& C_A^2 \left(-\frac{2573}{5184} + \frac{11}{24}\,z(2)\right) + 
       C_F^2 \left(\frac{35}{648} + \frac{z(3)}{3}\right)  
  \Bigg];
\\
\bar{G}_{X 32}&=&
\frac{C_F}{\pi^3}
 \Bigg[ \left( \frac{229}{1296} - \frac{z(2)}{36} \right) \,n_f^2 +
    C_F\,C_A\left(\frac{2807}{1296} + \frac{1183}{288}\,z(2)
    - \frac{11}{3}\,z(3)+\frac{5}{4}\,z(4)\right) +\nonumber\\ 
&+&        C_F\,n_f\left(-\frac{401}{1296} - \frac{121}{144}\,z(2)
     + \frac{5}{12}\,z(3)    \right) + 
        C_F^2 \left(\frac{811}{288} + \frac{13}{36}\,z(2)
    - \frac{31}{12}\,z(3)    - \frac{11}{4}\,z(4)    \right) + \nonumber\\ 
&+&        C_A\,n_f \left(-\frac{10115}{5184} + \frac{29}{72}\,z(2)
    - \frac{z(3)}{24}  \right) + 
        C_A^2 \left(\frac{6217}{1296} - \frac{29}{24}\,z(2)
    + \frac{77}{48}\,z(3)  - \frac{11}{16}\,z(4)    \right) 
  \Bigg].
\end{eqnarray}
The coefficient of the single logarithm at $O(\alpha)$, $\bar{G}_{X11}$, 
is the same as in the distribution in the hadron mass squared $t$, 
while the higher-order coefficients are different.
Therefore there is not a simple relation between these two spectra beyond 
leading order from two loops on, as was instead guessed in \cite{mesbagliato}.

\subsection{Study of the $BLM$ scheme}

In this section we compare the exact second-order coefficients
in the exponent of the electron form factor $\bar{G}_X$ with those ones
obtained by the $BLM$ ansatz:
\bea
\bar{G}_{X22} &=& 
\frac{2053}{648\,\pi^2}+\frac{1}{54}-\frac{23}{108\,\pi ^2}\,n_f
\,=\,
0.339525\, -\, 0.0215777 \, n_f
\\
\hat{G}_{X22} &=& 
-\,\frac{23}{108\,\pi ^2}\,
\left(n_f\,-\,\frac{33}{2}\right)
\,=\,
0.356031\, -\, 0.0215777 \, n_f
\\
\bar{G}_{X21} &=& 
-\,\frac{11\,z(3)}{3\,\pi ^2}\,+\,\frac{1405}{54\,\pi ^2}\,-\,\frac{55}{81}\,-\,
\left(\frac{73}{54\,\pi ^2}\,-\,\frac{1}{27}\right)\,n_f
\,=\,
1.51064\, -\, 0.0999342 \, n_f
\\
\hat{G}_{X21} &=& 
-\,\left(\frac{73}{54\,\pi ^2}\,-\,\frac{1}{27}\right)
\left(n_f\,-\,\frac{33}{2}\right)
\,=\,
1.64891\, -\, 0.0999342 \, n_f
\eea
As in the previous case, this approximation works within 
$O(10\%)$, i.e. rather well.

\subsection{Non-minimal scheme}

As discussed in detail in \cite{noi2}, 
for phenomenological applications it is convenient to define a non-minimal
scheme which involves the integration of the universal form factor 
$\Sigma(u;\alpha)$ only in the physical region $0<u\le 1$.
Let us introduce also for this scheme an over-all factor equal to the spectrum 
in lowest order:
\begin{equation}
\frac{1}{2\,\Gamma}\,\frac{d\Gamma}{d\bar{x}} \, = \, 
( 1 \, - \, \bar{x} )^2 \, ( 1 \, + \, 2 \, \bar{x} )
\, \Big[
C_X(\alpha) \, \Sigma_X(\bar{x};\,\alpha) \, + \, d_X(\bar{x};\,\alpha)
\Big].
\end{equation}
We choose to define the coefficient function as:
\begin{equation}
C_X(\alpha) \,  \equiv \,  \frac{1}{2} \, \int_0^1 dw \, C_L(0,w;\,\alpha).
\end{equation}
One immediately obtains for the first-order correction:
\begin{equation}
C_X^{(1)} \, = \, - \, \frac{C_F}{\pi} \frac{127}{72} \, = \, - \, 0.748618.
\end{equation}
The correction to the coefficient function is
--- as in all the cases we have considered ---
negative. For $\alpha(m_b)=0.22$ the correction is $\approx - 16.5\%$, i.e.
it is basically half of that in the minimal scheme.

The effective electron form factor can be defined as:
\begin{eqnarray}
\Sigma_X(\bar{x};\,\alpha) &=&
\frac{\int_0^1 dw \,C_L(0,w;\,\alpha)
\,\tilde{\Sigma}\left[ \bar{x}/w; \, \alpha(w\,m_b) \right]}
   { \int_0^1 dw \,C_L(0,w;\,\alpha) }
\nonumber\\
&=& \frac{
\int_0^{\bar{x}} dw \,C_L(0,w;\,\alpha)
\, + \,
\int_{\bar{x}}^1 dw \,C_L(0,w;\,\alpha)
\,\Sigma\left[ \bar{x}/w; \, \alpha(w\,m_b) \right]
}{ \int_0^1 dw \,C_L(0,w;\,\alpha) },
\end{eqnarray}
where $\alpha=\alpha(m_b)$. In order to simplify the definition
as much as possible,
we have integrated down to $w=0$, replacing the form factor $\Sigma[u;\,\alpha]$ 
with its extension $\tilde{\Sigma}[u;\,\alpha]$.
The normalization is the same as in the minimal scheme:
\begin{equation}
\Sigma_X(1;\,\alpha)\, = \, 1.
\end{equation}
The perturbative expansion of the form factor is:
\begin{equation}
\Sigma_X(\bar{x};\,\alpha) \,=\, 1 \, + \, \alpha \, \Sigma_X^{(1)}(\bar{x})
\, + \, \alpha^2 \, \Sigma_X^{(2)}(\bar{x}) \, + \, O(\alpha^3).
\end{equation}
By replacing the explicit expressions for the coefficient function $C_L$ and
the form factor $\Sigma$, expanding in $\alpha$ and performing the integration, 
we obtain:
\begin{equation}
\label{SigmaXnonmin}
\Sigma_X^{(1)}(\bar{x})\, = \, \frac{C_F}{\pi}
\left(
\, - \, \frac{1}{2} L_{\bar{x}}^2 \, + \, \frac{31}{12} L_{\bar{x}} \, 
- \, \frac{143}{72} \, + \, \frac{27}{8}\,\bar{x}^2 
\, - \, \frac{25}{18}\,\bar{x}^3
\right).
\end{equation}
In general, the form factor contains constant terms for $\bar{x}\rightarrow 0$
as well as vanishing terms in the same limit. The former modify the coefficient
function while the latter modify the remainder function with respect to the
values in the minimal scheme.

The remainder function reads in this scheme: 
\begin{eqnarray}
\tilde{d}_X^{(1)}(\bar{x}) &=& \frac{C_F}{\pi} \, 
\Bigg\{
\, \frac{29}{12} \, - \, \frac{27}{8} \, \bar{x}^2 \, + \, \frac{25}{18} \, \bar{x}^3
\, - \, \frac{41}{36\,(1 - \bar{x})}
\, - \, \frac{23}{ 18\,\left( 1 + 2\,\bar{x} \right) }
\, + \, \log\bar{x} \, \log(1-\bar{x}) + {\rm Li}_2(\bar{x}) 
\, + \,\nonumber\\
&&~~~~ + \,\left[ \frac{23}
      {12} - \frac{41}
      {36\,(1-\bar{x})^2} + 
     \frac{13}{54\,(1-\bar{x})} - 
     \frac{55}
      {54\,\left( 1 + 2\,\bar{x}
          \right) } \right] \,
   \log \bar{x} \,
\Bigg\}.
\end{eqnarray}
This remainder function also vanishes for $\bar{x}\rightarrow 0$
and differs from the one in the minimal scheme only for the
subtraction of the infinitesimal terms in (\ref{SigmaXnonmin}).

The coefficients of the infrared logarithms $G_{Xij}$ in the exponent
of the form factor $G_X$ are the same as those in the minimal scheme 
computed in the previous section:
\begin{equation}
 G_{X ij} \, = \, \bar{G}_{X ij}~~~~~~~~~~~~~~~~~~~~~~~~~~~~~{\rm for}~j\ge 1.
\end{equation}
In the limit $\bar{x}\rightarrow 0$ (i.e. neglecting $O(\bar{x})$ terms), 
$G_X$ also contains  the constant terms:
\begin{eqnarray}
G_{X10} &=& 
- \, \frac{19}{36} \,A_1 \, + \,  
  \frac{5}{6} \,\left( B_1 + D_1 \right);
\\
G_{X20} &=& 
- \, \frac{19}{36} \,A_2 
\, - \, \frac{65}{72} \,A_1\,\beta_0 
\, + \, \frac{13}{72} \, {\left( B_1 + D_1 \right) }^2 
\, + \, \frac{19}{36} \,\beta_0 \, \left( 3\,B_1 + 2\,D_1 \right) 
\, + \, \frac{5}{6}\,\left( B_2 + D_2 \right)  
\, + \nonumber\\
&+& A_1\,\left( B_1 + D_1 \right) \, \left( - \frac{25}{54} 
\, + \, \frac{5}{6}\,z(2) \right)  
\, + \, \left( B_1 + D_1 \right) \, \frac{C_F }{\pi }\,\left( \frac{583}{216} - 
       2\,z(3) \right) 
\, + \nonumber\\
&+& 
  {A_1}^2\,
   \left( \frac{905}{2592} - 
     \frac{19}{36}\,z(2) + \frac{5}{6}\,z(3)
     \right)  + A_1\, \frac{C_F}{\pi}\,
     \left( - \frac{7337}{1296} + 
       \frac{5}{3}\,z(3) + 3\,z(4) \right). 
\end{eqnarray}
Explicitly:
\begin{eqnarray}
G_{X10} &=& - \, \frac{C_F}{\pi} \, \frac{143}{72};
\\
G_{X20} &=& 
\frac{C_F}{{\pi }^2}\,
\Bigg[ 
\, n_f\,\left( \frac{1507}{1296} - 
         \frac{5}{36}\,z(2) \right)  + 
      C_A\,
       \left( - \frac{2101}{324}
             + \frac{13}{9}\,z(2) + 
         \frac{5}{24}\,z(3) \right)  
\, + \nonumber\\
&& ~~~~ + \, 
      C_F\,
       \left( - \frac{90725}{10368}
             - \frac{49}{36}\,z(2) + 
         \frac{19}{4}\,z(3) + 3\,z(4) \right) 
\Bigg]. 
\end{eqnarray}
The relations between the coefficients functions in the two schemes
have the same form as those for the $\hat{p}_+$ spectrum given in
the previous section:
\begin{eqnarray}
C_X^{(1)} &=& \bar{C}_X^{(1)} - G_{X10};
\\
C_X^{(2)} &=& \bar{C}_X^{(2)} \, - \, \bar{C}_X^{(1)} \, G_{X10}
\, + \, \frac{1}{2} \, G_{X10}^2 \, - \, G_{X20}.
\end{eqnarray}
The first of the above equations can be directly verified by inserting
the first-order quantities. 

Long-distance phenomena (large logarithms, Fermi motion, hadronization, etc.) 
are expected to have a larger effect in the distributions
in the variables $q=t,u$ or $p$ than in the electron energy spectrum
\footnote{The $u$ spectrum have been studied in \cite{noi2}.}.
That is because the former variables have a tree-level distribution
consisting of a peak in zero:
\begin{equation}
\label{peak}
~~~~~~~~~~~~~~~~~~~~~~~~~~~~~~~~~~~~~~~~~~~~~
\frac{1}{\Gamma}\,\frac{d\Gamma}{dq} \, = \, \delta(q) \, + \, O(\alpha)
~~~~~~~~~~~~~~~~~~~~~~~~~~~~~~~~~~~~~~~~~~~(q\,=\,u,\,t,\,p),
\end{equation}
while the electron spectrum has a broader distribution with a maximum
in $\bar{x}=0$:
\begin{equation}
\label{smeared}
\frac{1}{2 \, \Gamma} \, \frac{d\Gamma}{d\bar{x}} \, = 
\, (1 \, - \, \bar{x})^2 \, (1 \, + \, 2 \, \bar{x}) \, + \, O(\alpha).
\end{equation}
Long-distance phenomena always have a smearing effect, which is more
prounanced in a starting distribution of the form (\ref{peak}) than of the
form (\ref{smeared}). 
Technically, that is reflected by the fact that
long-distance effects in the electron spectrum are tempered by the 
presence of the partially-integrated form factor $\Sigma$ instead of 
the differential one $\sigma$.
We therefore expect the electron spectrum to be a less sensitive
quantity to threshold effects.

\section{The $BLM$ scheme for other spectra} 
\label{nuovasec}

The accuracy of the $BLM$ ansatz can also be studied
by comparing its predictions with the exact results obtained
for the following distributions treated in \cite{noi} and 
in \cite{noi2}: 
\begin{enumerate}
\item
hadron mass distribution in the radiative decay (\ref{startrd})
(or, equivalently, semileptonic distributions not integrated
over the hadron energy). Our results for the $QCD$ coefficients
and the $BLM$ estimates read respectively:
\bea
      G_{22} &=& 
\frac{95}{72\, \pi ^2}\,+\,\frac{1}{54}-\frac{13}{108\, \pi ^2}\,n_f
\,=\,
0.152206\, -\, 0.0121961 \, n_f;
\\
\hat{G}_{22} &=& 
-\frac{13}{108\, \pi ^2}\,
\left(n_f\,-\,\frac{33}{2}\right)
\,=\,
0.201235\, -\, 0.0121961 \, n_f;
\\
      G_{21} &=& 
\,-\,\frac{
   z(3)}{9\, \pi ^2}\,+\,\frac{917}{72\, \pi ^2}\,-\,\frac{35}{54}
\,-\,\left(\frac{85}{108\,\pi ^2}\,-\,\frac{1}{27}\right)\,n_f
\,=\,
0.628757\, -\, 0.0427065 \, n_f;
\\
\hat{G}_{21} &=& 
-\left(\frac{85}{108\,\pi ^2}\,-\,\frac{1}{27}\right)\,
\left(n_f\,-\,\frac{33}{2}\right)
\,=\,
0.704657\, -\, 0.0427065 \, n_f.
\eea
The estimate of the non-abelian contribution is accurate
with $O(25\%)$;
\item
distribution in the variable $u$ defined in the introduction
(basically the hadron mass normalized to the hadron energy).
We obtain:
\bea
G_{U22} &=& 
-\,\frac{5}{24 \,\pi ^2}\,+\,\frac{1}{54}\,-\,\frac{1}{36 \,\pi ^2}\,n_f
\,=\,
-\,0.00259006 \,-\, 0.00281448 \, n_f;
\\
\hat{G}_{U22} &=& 
-\,\frac{1}{36 \,\pi ^2}\,
\left(n_f\,-\,\frac{33}{2}\right)
\,=\,
0.0464389 \,-\, 0.00281448 \, n_f;
\\
G_{U21} &=& 
-\,\frac{z(3)}{9 \,\pi ^2}\,+\,\frac{217}{12 \,\pi^2}\,-\,\frac{35}{54}
\,-\,\left(\frac{10}{9 \,\pi ^2}\,-\,\frac{1}{27}\right)\,n_f
\,=\,
1.17054 \,-\, 0.0755421 \, n_f;
\\
\hat{G}_{U21} &=& 
-\left(\frac{10}{9 \,\pi ^2}\,-\,\frac{1}{27}\right)\,
\left(n_f\,-\,\frac{33}{2}\right)
\,=\,
1.24644 \,-\, 0.0755421 \, n_f;
\eea
The agreement is good for the $G_{U21}$ coefficient, within
$10\%$, but not good for the $G_{U22}$ coefficient, which is however 
very small because of some accidental cancellation;
\item
hadron mass spectrum in the semileptonic case.
We obtain:
\bea
G_{T22} &=& 
\frac{1057}{216\,\pi^2}+\frac{1}{54}
\,-\,\frac{11}{36 \,\pi^2}\,n_f \,=\,
0.514336\, - \,0.0309593 \, n_f
\\
\hat{G}_{T22} &=& 
-\,\frac{11}{36 \,\pi^2}\,
\left(n_f\,-\,\frac{33}{2}\right)
\,=\,
0.510828\, - \,0.0309593 \, n_f;
\\
G_{T21} &=& 
-\,\frac{65\,z(3)}{9\, \pi^2}+\frac{12431}{648 \,\pi ^2}-\frac{55}{81}
\,-\,\left(\frac{83}{108 \,\pi ^2}\,-\,\frac{1}{27}\right)\,n_f\,=\,
0.385075\, - \,0.0408302 \, n_f;
\\
\hat{G}_{T21} &=& 
-\left(\frac{83}{108 \,\pi ^2}\,-\,\frac{1}{27}\right)\,
\left(n_f\,-\,\frac{33}{2}\right)
\,=\,
0.673698\, -\, 0.0408302 \, n_f.
\eea
The estimate of $G_{T22}$ is very accurate, while for 
$G_{T21}$ it is not.
\end{enumerate}
The over-all picture is that the $BLM$ ansatz 
offers an estimate of the resummation coefficients  
with an acceptable relative error most of the times.
There are ``exceptional cases'', when the coefficients are very small 
because of some accidental cancellation,  which cannot 
be accounted for by the general $BLM$ ansatz.

\section{Conclusions}
\label{conclude}

In this paper we have resummed to next-to-leading order the distributions in the light-cone
momentum $p_+ \, = \, E_X \, - \, |\vec{p}_X|$ and in the charged lepton energy $E_e$ 
in the semileptonic decays (\ref{startsl}), where $E_X$ and $\vec{p}_X$ are the total energy 
and three-momentum of the final hadron state $X_u$. 
These spectra have a different threshold structure with respect
to the hadron energy distribution in (\ref{startsl}) or the photon spectrum 
in (\ref{startrd}), because they involve integration over $E_X$, 
i.e. over the hard scale $Q \, = \, 2E_X$ of the hadronic subprocess in (\ref{startsl}).
We have also presented a qualitative discussion of specific aspects of the
long-distance effects in the electron energy spectrum: they should be 
suppressed with respect to the distribution in the hadron mass $m_X$ 
or in $p_+$ because of a smearing effect of the Born three-body 
kinematics.

We have explicitly checked our resummation formula for the $p_+$ form factor
by expanding it to $O(\alpha^2)$, picking up the $n_f$ contributions and comparing them 
with the $O(\alpha^2 \, n_f)$ corrections computed with Feynman diagrams.
This is a highly non-trivial check at the two-loop level of the resummation formula
for the triple differential distribution and makes us more confident of our formalism.

We have also discussed the validity of the so-called Brodsky-Lepage-Mackanzie 
($BLM$) scheme for the estimate of the second-order corrections to the form
factors for various spectra.
The conclusions are that this scheme offers in general a rather good 
approximation: it exactly reproduces the coefficients of the
$\alpha^2 \, \log^3$ terms and gives good numerical estimates of the
$\alpha^2 \, \log^2$ and $\alpha^2 \, \log$ coefficients.
In cases in which the $QCD$ coefficients are anomaously small because
of some accidental cancellation, the $BLM$ scheme is not accurate, 
as in these cases the $n_f$-dependent and the $n_f$-independent
contributions presumably cancel to a different degree.
It would be interesting to extend the $BLM$ analysis to the third-order
coefficients which have been computed by us for all the distributions
treated.
Let us remark that the exponentiation property does not follow from the
$BLM$ ansata and it has to be imposed externally.

By expanding the resummed formulas up to $O(\alpha^3)$, we have computed the
coefficients $\bar{G}_{P ij}$ and $\bar{G}_{X ij}$ of the infrared logarithms 
$\log m_b/p_+$ and $\log 1/\bar{x}$ in the exponent of the effective 
form factors $G_P$ and $G_X$ respectively.
These coefficients constitute a true prediciton of our resummation schme
and can be checked with second and third order computations as soon as
they become available. 
These coefficients can also be used in phenomenological studies which do not
implement the full resummed formulas.
For both distributions we have defined (standard) minimal factorization schemes
and also non-minimal schemes which can be easily implemented in phenomenological 
analyses.
As in our previous work \cite{noi2}, we find that the perturbative expansion
seems to have better convergence properties in the modified schemes than in
the minimal one.

Analytic expressions for double distributions or for single distributions
with some kinematical cut can be obtained in similar way as we have made.
This work completes however our study of the general factorization-resummation 
properties of semileptonic decay spectra \cite{me,noi,noi2}: 
for each distribution considered we have computed the coefficient function and 
the remainder function in NLO and the (effective) form factor in NNLO. 
The next step is a phenomenological analysis of the experimental data 
on semileptonic and radiative decays by using a model for the QCD form 
factor such as that formulated in \cite{model}, combined eventually 
with non-perturbative components not calculable in perturbatve QCD.

\vskip 0.3 truecm

\centerline{\bf Note added}

\vskip 0.2 truecm

\noindent
A few days after the first version of this paper was put on the
archive, another work on resummation in semi-leptonic decays has 
also appeared \cite{gardifinal}.
In this paper, expressions of the second-order coefficients 
of the $p_+$ form factor are derived, which are in complete
agreement with ours.

\end{document}